\documentclass[twocolumn,showpacs,preprintnumbers,prb,fleqn,superscriptaddress]{revtex4-1}
\usepackage{graphicx}
\usepackage{dcolumn}
\usepackage{bm}
\usepackage{natbib}

\newcommand{\arctanh}{\mathop{\mathrm{arctanh}}\nolimits}
\renewcommand{\Im}{\mathop{\mathrm{Im}}\nolimits}
\renewcommand{\vec}[1]{\mathbf{#1}}
\renewcommand{\Re}{\mathop{\mathrm{Re}}\nolimits}

\begin{document}

\title{Electronic excitations and electron-phonon coupling in bulk graphite through Raman scattering in high magnetic fields}

\author{P. Kossacki}
\affiliation{LNCMI, UPR 3228, CNRS-UJF-UPS-INSA, 38042 Grenoble,
France.} \affiliation{Institute of Experimental Physics, Faculty
of Physics, University of Warsaw, Poland.}
\author{C. Faugeras}\email{clement.faugeras@lnmci.cnrs.fr}\affiliation{LNCMI, UPR 3228, CNRS-UJF-UPS-INSA, 38042 Grenoble, France.}

\author{M. K\"{u}hne} \affiliation{LNCMI, UPR 3228, CNRS-UJF-UPS-INSA, 38042 Grenoble, France.}
\author{M. Orlita}
\affiliation{LNCMI, UPR 3228, CNRS-UJF-UPS-INSA, 38042 Grenoble,
France.}
\author{A.A.L. Nicolet}
\affiliation{LNCMI, UPR 3228, CNRS-UJF-UPS-INSA, 38042 Grenoble,
France.}
\author{J.M. Schneider}
\affiliation{LNCMI, UPR 3228, CNRS-UJF-UPS-INSA, 38042 Grenoble,
France.}
\author{D. M. Basko}
\affiliation{Universit\'e Grenoble 1/CNRS, LPMMC UMR 5493, 25 rue des
Martyrs, 38042 Grenoble, France}
\author{Yu. I. Latyshev}
\affiliation{Institute of Radio-Engineering and Electronics RAS,
Mokhovaya 11-7, 101999, Moscow, Russia}
\author{M. Potemski}
\affiliation{LNCMI, UPR 3228, CNRS-UJF-UPS-INSA, 38042 Grenoble,
France.}
\date{\today }

\begin{abstract}
We use polarized magneto-Raman scattering to study purely
electronic excitations and the electron-phonon coupling in bulk
graphite. At a temperature of 4.2~K and in magnetic fields up to
28~T we observe $K$-point electronic excitations involving Landau
bands with $\Delta |n|=0$ and with $\Delta |n|=\pm2$ that can be
selected by controlling the angular momentum of the excitation
laser and of the scattered light. The magneto-phonon effect
involving the $E_{2g}$ optical phonon and $K$-point inter Landau
bands electronic excitations with $\Delta |n|=\pm1$ is revealed
and analyzed within a model taking into account the full $k_z$
dispersion. These polarization resolved results are explained in
the frame of the Slonczewski-Weiss-McClure (SWM) model which
directly allows to quantify the electron-hole asymmetry.
\end{abstract}
 \maketitle

\section{Introduction}

Magneto-optical spectroscopy has been used to study the
graphene/graphite structures for long time, but mostly
transmission/reflectivity type of experiments (essentially in
far-infrared spectral range) have been explored so
far~\cite{Galt1956,Schroeder1968,Toy1977,Li2006,Sadowski06,Jiang07,Orlita08,Orlita08a,Henriksen2008,Orlita2009,Chuang2009,Orlita2010}.
Relevant for these systems, low energy electronic excitations can
be optionally probed with Raman scattering
methods~\cite{Faugeras2011}. These methods have been widely used
to investigate phonon resonances in different carbon
materials~\cite{Dresselhaus2010}, but advantages of combining them
with application of magnetic fields have been recognized only
recently~\cite{Ando07,Goerbig2007,Kashuba2009,Mucha-Kruczynski2010}.

Indeed, magneto-Raman scattering experiments provide a spectacular
demonstration of hybridization between optical phonon and
electronic excitations in epitaxial graphene~\cite{Faugeras2009}
and graphene locations on graphite
surface~\cite{Yan2010,Faugeras2011}. What is perhaps even more
important is that such experiments allow also to probe purely
electronic excitations in these systems~\cite{Faugeras2011}. When
a magnetic field is applied across the layer, the electronic bands
condense into Landau levels whose characteristic energy ladders
(fan charts) reflect the specific dispersion relation of
electronic states in the absence of the magnetic field.
Magneto-Raman scattering can be used to trace selected inter
Landau level excitations. This method, with all its advantages of
a visible optics technique (spatial focusing, polarization
resolved measurements), appears now as a valuable option for
Landau level spectroscopy to study other sp$^2$-bonded carbon
allotropes, such as bulk graphite, investigated in this work.

In spite of recent efforts~\cite{Garcia09}, the magneto-Raman
scattering response of bulk graphite has not been so far clearly
identified in experiments. Theoretical studies addressing this
specific problem seem to be also missing. Relevant for our results
are, however, predictions for a bilayer
graphene~\cite{Mucha-Kruczynski2010} which is the basic unit in
construction the Bernal-stacking of graphite.

As a layered material, graphite is strongly anisotropic but still
represents a three dimensional electronic system. In this respect,
graphite is very different from its layer-components: purely
two-dimensional monolayer or bilayer graphene. Electrons can
easily move within the graphitic layers, but electronic states
also remain dispersive in the direction across the layers. When a
magnetic field is applied across the layers, we deal with Landau
bands in graphite in contrast to discrete Landau levels in purely
2D systems (monolayer or bilayer graphene). Band structure of
graphite is commonly described using the SWM
model~\cite{Slonczewski58, McClure56}. There is a general
consensus on the validity of this tight binding approach. However,
it implies up to seven hoping-integral parameters and the exact
values of some of them remain controversial and this is despite
very long history of the research on graphite.

Our magneto-Raman scattering experiments on graphite reveal a
series of features which we attribute to electronic excitations
between Landau bands in this material. The magnetic field
evolution of the E$_{2g}$ phonon excitation is also investigated.
The observed electronic and phonon excitations are shown to follow
the characteristic selection rules defined by the appropriate
transfer of angular momentum between the incoming and outgoing,
circularly-polarized photons. Polarization-resolved experiments
together with appropriate modelling of electronic bands allow us
to conclude about the relevant electron-hole asymmetry of the
graphitic bands. Although our experiments are mostly sensitive to
electronic states in the vicinity of the particular K-point of the
graphite Brillouin zone, the three dimensional nature of this
material is clearly reflected in the spectral shape of the
observed electronic excitations and in the character of the
investigated magneto-phonon resonances. Notably the E$_{2g}$
phonon of graphite is shown to hybridize with a quasi-continuous
spectrum of inter Landau band excitations, in contrast to similar
effect in graphene involving the discrete LL transitions.

The paper is organized as follows: section II describes the
experimental set-up, while the electronic properties of bulk
graphite and the Raman scattering selection rules are presented in
section III. Experimental results obtained in the co- and
crossed-circular polarization configuration are presented in
section IV, together with the magneto phonon effect of bulk
graphite and experiments performed at room temperature. We finally
present our conclusions.

\section{Experiment}

The polarized Raman scattering response of bulk graphite at low
temperature and in high magnetic fields, was measured with a home
made confocal micro-Raman scattering set-up. A monolithic
miniaturized optical table, made of titanium, has been designed to
operate in a helium exchange gas at T$=4.2$K and in high magnetic
field environments. A mono-mode optical fiber with
$5\,\mu\mbox{m}$ core was used to bring the
$\lambda=784\,\mbox{nm}$ excitation from a Ti:Saphire laser to the
sample. The excitation beam is focused with lenses down to $\sim
1\,\mu\mbox{m}$ spot. Scattered light is then collected by a
multi-mode $200\,\mu\mbox{m}$ optical fiber before being analyzed
by a $500$~mm spectrometer equipped with a diffraction grating and
a nitrogen cooled CCD. Our miniaturized optical table can host
different optical filters that are used at liquid helium
temperature and in high magnetic fields. Band-pass filters (laser
line and notch filters) are used first to clean the laser coming
out of the mono-mode fiber and to filter the elastically scattered
laser before the collection optical fiber. They impose a cut-off
energy of $\sim 350\,\mbox{cm}^{-1}$ from the laser line. A set of
quarter wave plates and linear polarizers, placed close to the
sample, are then used to circularly polarize the excitation beam
and the collected signal to achieve both co- and crossed-circular
polarization configurations. The two different co-circular
($\sigma^-/\sigma^-$ and $\sigma^+/\sigma^+$) or cross-circular
($\sigma^-/\sigma^+$ and $\sigma^+/\sigma^-$) configurations were
obtained by changing the direction of the magnetic field with
respect to the light propagation direction. Optical power on the
sample was fixed to $5\,\mbox{mW}$. The sample is mounted on a set
of translation piezzo stages which allow to move the sample under
the laser spot with a sub-micrometer resolution.

The surface of natural graphite is known to be inhomogeneous and
graphene flakes decoupled from the surface can be found. They have
been evidenced by low temperature STM experiments~\cite{Li2009},
studied by an EPR-like technique in magnetic
fields~\cite{Neugebauer2009} and mapped with micro-Raman
scattering measurements in high magnetic fields revealing their
graphene-like electronic excitation spectrum~\cite{Faugeras2011}
and the associated magneto-phonon effect~\cite{Yan2010,
Faugeras2011}. To unravel the electronic properties of bulk
graphite, we place the laser spot outside of these decoupled
graphene flakes and measure the magnetic field evolution of the
Raman scattering spectrum. Similar results were obtained using two
different sources of natural graphite and Highly Oriented
Pyrolitic Graphite (HOPG) and in the following, we will only
discuss the case of natural graphite which shows, in our
experiment, a slightly higher signal level.

\section{Theoretical outline}

\subsection{Graphite band structure}

A conventional description of the band structure of bulk graphite
and its evolution with the magnetic field relies on the SWM model
with its seven $\gamma_0,\ldots,\gamma_5,\Delta$ tight binding
parameters~\cite{Slonczewski58, McClure56,Nakao1976}. This model
has been used to describe most of previous data obtained from
magneto-transport~\cite{Soule1958,Soule1964,Woollam1970,Schneider2009},
infrared magneto-reflectivity~\cite{Schroeder1968,Toy1977,
Li2006}, and
magneto-transmission~\cite{Orlita08a,Orlita2009,Chuang2009}
experiments. It predicts the existence of massive carriers near
the $K$ point with a parabolic in-plane dispersion and of massless
carriers near the $H$ point with a linear in-plane dispersion. The
Fermi energy is $\sim 15\,\mbox{meV}$ and, under an applied
magnetic field, Landau bands are formed with a continuous
dispersion along $k_z$, from equally spaced and linear in $B$
Landau levels at the $K$ point to non-equally spaced and $\sqrt B$
evolving Landau levels at the $H$ point~\cite{Nakao1976}. Even
though there is still no consensus concerning the precise values
of the SWM parameters, mainly because of the different energy
ranges probed in different experiments and because of the lack of
polarization resolved measurements that would reveal unambiguously
the effect of electron-hole asymmetry, the validity of the SWM
model is generally accepted.

\begin{figure}
\includegraphics[width=0.65\linewidth,angle=0,clip]{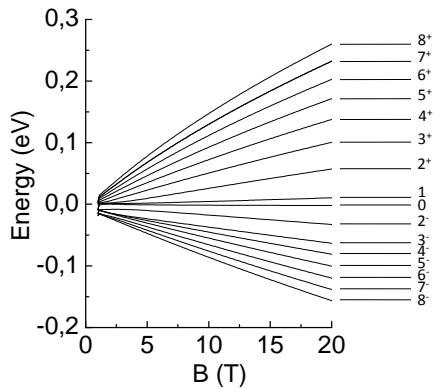}
\caption{\label{LandauLevels} Evolution of the Landau levels at
the K point of the graphite first Brillouin zone calculated within
the SWM model and Landau levels' indices.}
\end{figure}

The evolution in magnetic field of electronic levels in bulk
graphite has been calculated following the approach of
Nakao~\cite{Nakao1976} and Landau levels are labelled following
the bilayer graphene convention~\cite{Abergel2007} as sketched in
Fig.~\ref{LandauLevels}. The infinite order magnetic field
Hamiltonian was reduced to a $600 \times 600$ matrix before the
diagonalization procedure. Electronic excitations are labelled as
$L_{n,n'}$ where $n$ and $n'$ are the indices of the Landau levels
involved in the excitation $n\to{n}'$.

Instead of the full SWM model, it is often sufficient to use the
effective two-parameter model, which gives parabolic dispersion in
the plane with the slope of the parabolas depending on~$k_z$, the
wave vector measured in the units of the inverse inter-layer
spacing ($k_z=0$ at the $K$~point, and $k_z=\pi/2$ at the~$H$
point). This model is obtained by (i)~neglecting all SWM
parameters except two, $\gamma_0$~and~$\gamma_1$, the intralayer
and the interlayer nearest-neighbor hopping integrals,
respectively, and (ii)~projecting the resulting $k_z$-dependent
$4\times{4}$ Hamiltonian on the two low-energy bands. At each
value of~$k_z$, the Hamiltonian is identical to that of a graphene
bilayer determined by the effective parameters $\gamma_0$ and
$\gamma_1^*=2\gamma_1\cos k_z$. Since at the $K$ point $k_z=0$,
the corresponding $\gamma_1^*$ is twice enhanced with respect to
$\gamma_1$ describing the real graphene
bilayer~\cite{Partoens2007,Koshino2008}. This effective
two-parameter parabolic model has proven to bring a fair frame to
describe magneto-optical experiments~\cite{Orlita2009,
Chuang2009}.

Still, the effective two-parameter parabolic model misses several effects, such
as the in-plane trigonal warping, described by the $\gamma_3$
parameter of the full SWM model, as well as the electron-hole
asymmetry.
The latter is contributed by all remaining parameters
($\gamma_2,\gamma_4,\gamma_5,\Delta$). To simplify the analysis,
one can notice that at $k_z=0$, the $\gamma_2,\gamma_5,\Delta$
do not enter independently; eigenstates and transition energies
depend only on their combination
$\Delta_{eff}\equiv\Delta+2\gamma_5-2\gamma_2$, as can be seen
from Eqs.~(\ref{bilayerHam=}),~(\ref{bilayerparams=}) of the
Appendix.
This combination determines the asymmetry in the positions of
the split-off bands exactly at the $K$~point. For non-zero values
of the in-plane wave vector, the electron-hole asymmetry is
determined by both $\gamma_4$ and $\Delta_{eff}$, and it is not
easy to separate the effects of the two.
Throughout the paper, we use a set of SWM parameters derived
from our
polarization-resolved magneto-Raman scattering experiments. We
will detail in the following sections how these parameters are
determined, essentially from the crossed circular polarization
configuration measurements which allow to directly quantify the
electron-hole asymmetry.

\subsection{Selection rules for Raman scattering}
\label{sec:selectionrules}

Generally, the Raman scattering selection rules for graphite can
be deduced from its symmetry group, $D_{6h}$. However, the
low-energy electronic Hamiltonian, as derived from the SWM model,
is dominated by terms which have a higher symmetry. As a
consequence, the electronic dispersion is almost isotropic with
respect to continuous rotations of electronic momentum around the
$H-K-H$ line. This isotropy is broken only by the trigonal warping
term, governed by the $\gamma_3$~parameter of the SWM model. This
term restricts the symmetry to the momentum rotations by
$\pm{2}\pi/3$. As a result, the quantum number~$m_z$, associated
with the rotational symmetry (angular momentum), which could
assume all integer values from $-\infty$ to $\infty$ in the case
of the continuous rotational symmetry, has only three distinct
values for the three-fold symmetry: all $m_z$'s differing by a
multiple of~3, become equivalent. However, since the trigonal
warping is small at energies, relevant for the present work, all
transitions which are allowed by $D_{6h}$, can be further
classified into \emph{strongly allowed} (those which are allowed
even when $\gamma_3=0$, thus surviving the continuous rotational
symmetry), and \emph{weakly allowed} (those which require a
non-zero trigonal warping). The wording "strongly" and "weakly"
allowed does not, however, refer to the apparent intensity of the
excitation.

An example of a weakly allowed process is the Raman process
responsible for the $G$~band feature. Indeed, trigonal warping is
necessary for the appearance of the $G$~band feature, as was shown
for monolayer graphene in Refs.~\onlinecite{Basko2008,Basko2009};
in bilayer graphene and graphite the situation is analogous. The
reason is that the circularly polarized photons carry an angular
momentum $m_z=\pm{1}$, and out of two degenerate phonon modes, one
can also choose two linear combinations carrying angular momentum
$m_z=\pm{1}$. Then if a $\sigma^+$ photon ($m_z=-1$) is absorbed
by the sample, the only way to conserve angular momentum is to
emit a $\sigma^-$ photon ($m_z=+1$) and a phonon with $m_z=+1$.
Then the total change in the angular momentum is $+3$, which is
equivalent to 0, thanks to trigonal warping. Thus, the $G$~band
feature is observed in the cross-circular polarization
configuration.

An inter-Landau-level excitation $n^-\to(n+m_z)^+$ carries an
angular momentum $m_z$. This immediately designates the
transitions with $\Delta|n|=0$ and $\Delta|n|=\pm{2}$ as the
strongly allowed processes in the co-circular and cross-circular
polarization configurations, respectively. As in the case of a
graphene monolayer, the strongly allowed coupling for the
magneto-phonon effect in a graphene bilayer~\cite{Ando2007b}
involves the optical phonon and electronic excitations with
$\Delta |n|=\pm 1$ (optical-like excitations). These optical-like
excitations are not expected to be Raman active without trigonal
warping. Nevertheless, if $\gamma_3$ is included then the trigonal
warping is expected to weakly allow these excitations. For
instance, an excitation $L_{n^-,(n+1)^+}$ with $\Delta |n|=+1$ can
be observed in Raman scattering experiments, because of the
trigonal warping induced mixing of the levels $n^-$ and $(n+3)^-$,
through the excitation $L_{(n+3)^-,(n+1)^+}$ which is strongly
allowed in Raman scattering. Interesting is the fact that
$\Delta|n|=\pm1$ are expected to be observed in the same crossed
circular polarization as $\Delta|n|=\mp2$ excitations. A signature
of such transitions has indeed been seen in the present
experiment.

\section{Experimental results and discussion}

\subsection{Co-circular configuration}
\label{cocircular}

\begin{figure}
\includegraphics[width=1\linewidth,angle=0,clip]{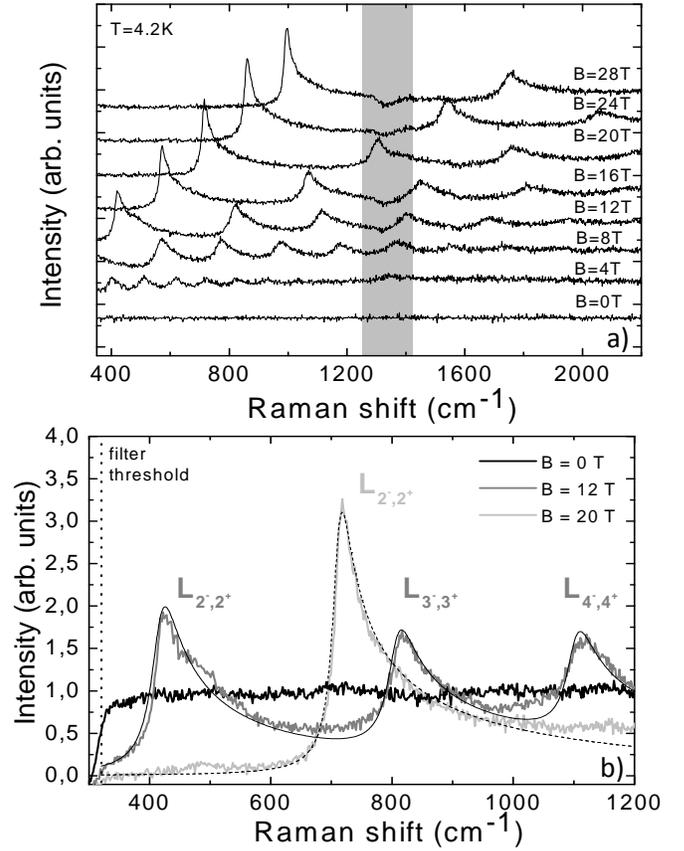}
\caption{\label{FigCoSpectre} a) Representative Raman scattering
spectra with subtracted $B=0$~T spectrum in the
$\sigma^-/\sigma^-$ configuration showing many asymmetric magnetic
field dependent features. Spectra are shifted for clarity. The
gray region indicates an energy range where a magnetic field
dependent background feature affects the normalized spectra. b)
Raw Raman scattering spectra to compare, at 3 different values of
the magnetic field, the scattered intensity. Solid and dashed
lines are calculated within the two-parameter parabolic model (see
main text for details). The vertical dotted line represents the
cut-off energy of the optical edge-filter used to remove the laser
line.}
\end{figure}

Fig.~\ref{FigCoSpectre}(a) shows representative Raman scattering
spectra in the co-circular polarization configuration, with the
$B=0$ spectrum subtracted. In this configuration, the $G$~band
feature is absent, as discussed in Sec.~\ref{sec:selectionrules}.
Starting from $B\sim{}2\,\mbox{T}$, many magnetic field dependent
features are observed, with energies increasing with the magnetic
field. The observed line shape is strongly asymmetric with a long
tail at high energies due the fact that the states involved in the
scattering process do not belong to discrete Landau levels, but to
Landau bands with a dispersion along~$k_z$. The width of these
features, about $50\,\mbox{cm}^{-1}$, is mostly determined by this
$k_z$~dispersion, rather by any scattering mechanism. To gain more
quantitative understanding of the line shape, we note that the
Raman matrix element does not depend on the energy of the
electronic states involved\cite{Mucha-Kruczynski2010}, so the
scattered intensity~$I(\omega)$ is simply proportional to the
joint density of states for excitations between Landau
levels~$n^-$ and~$n^+$:
\begin{equation}\label{IBpropto}
I_B(\omega)\propto\frac{1}{l_B^2}\int\limits_{0}^{\pi/2}dk_z
\sum_{n\geq{2}}
\delta\left(\omega-\epsilon_{n^+}+\epsilon_{n^-}\right),
\end{equation}
where $\epsilon_{n^\pm}$ are the Landau level energies, and the
magnetic length, $l_B=(eB/c)^{-1/2}$, keeps track of the Landau
level degeneracy. In the two-parameter parabolic model, the
Landau level energies are given by
\begin{equation}\label{epsilon=}
\epsilon_{n^\pm}= \pm\sqrt{n(n-1)}\,\frac{\Omega_B}{\cos{k}_z},
\quad
\Omega_B=\frac{\gamma_0^2}{\gamma_1}\left(\frac{3a}{2l_B}\right)^2,
\end{equation}
where $a$~is the in-plane distance between neighboring carbon
atoms. For $\epsilon_{n^\pm}$ given by Eq.~(\ref{epsilon=}),
Eq.~(\ref{IBpropto}) gives
\begin{equation}
\frac{I_B(\omega)}{I_{B=0}}=2\Omega_B
\sum_{n\geq{2}}\frac{\omega_n/\omega}{\sqrt{\omega^2-\omega_n^2}},
\end{equation}
where
$\omega_n\equiv\left.\epsilon_{n^+}-\epsilon_{n^-}\right|_{k_z=0}$.
This expression describes well the tail on the high-frequency side
of each peak, and has a square-root-type singularity at
$\omega\to\omega_n$. The singularity can be cut off by replacing
the $\delta$-function in Eq.~(\ref{IBpropto}) by a Lorentzian with
full width at half-maximum~$\Gamma$. This results in a somewhat
more complicated, but still explicit analytical expression for the
Raman spectrum:
\begin{eqnarray}
&&\frac{I_B(\omega)}{I_{B=0}}=\sum_{n\geq{2}}
\frac{2\Omega_B}{\omega_n}
\Im{f}\!\left(\frac{\omega-i\Gamma}{\omega_n}\right),
\label{IBIB0=}\\
&&f(x)=\frac{1}{2x}+\frac{2/\pi}{x\sqrt{x^2-1}}\, \arctanh
\sqrt{\frac{x+1}{x-1}}. \label{fx=}
\end{eqnarray}
In fact, the simple expression~(\ref{epsilon=}) does not reproduce
well the peak positions; the full SWM model taking into account
the electron-hole asymmetry is needed for this, as will be
discussed below. Still, if we take Eqs.
(\ref{IBIB0=}),~(\ref{fx=}) and substitute for $\omega_n$ the
actual peak positions, the resulting expression describes the
spectrum remarkably well, as shown in Fig.~2(b) for two values of
the magnetic field. We have taken $\Gamma=14\,\mbox{cm}^{-1}$ for
$B=12\,\mbox{T}$ and $\Gamma=20\,\mbox{cm}^{-1}$ for
$B=20\,\mbox{T}$. The discrepancy between Eq.~(\ref{IBIB0=}) and
the experimental spectrum around the minima between the peaks can
be improved by taking the full SWM result for $\epsilon_{n^\pm}$.

Fig.~\ref{FigCoSpectre}(b) shows that there exist an electronic
contribution to the $B=0$ Raman scattering spectrum of bulk
graphite, which is flat as a function of energy (between $300$ and
$1200$~cm$^{-1}$), but that can be identified by applying a
magnetic field. With the three Raman scattering spectra presented
in this figure, one can directly compare the amplitude of the
scattered light. When a magnetic field is applied, the energy
independent response from low energy electronic excitations
transforms into discrete features due to Landau quantization and
one can identify the "apparent background" signal in this
experiment with the scattered amplitude at energies lower than the
$L_{2^-,2^+}$ excitation, which is set to zero in
Fig.~\ref{FigCoSpectre}(b). In this particular experiment, we use
magnetic field to change the electronic excitation spectrum, but
one could expect a similar trend for gated graphene flakes.
Changing the gate voltage would result in a continuous change of
the $2E_F$ threshold in analogy to the tuning of the absorption
threshold as a function of the Fermi energy observed in infrared
absorption experiments~\cite{Li2009n}. One may also notice that
the amplitude of the scattered light between any two given peaks
L$_{n^-,n^+}$ and L$_{(n+1)^-,(n+1)^+}$ does not depend on the
magnetic field. This can be seen directly from
Eq.~(\ref{IBpropto}), which does not rely on
approximation~(\ref{epsilon=}): as long as Landau level energies
are proportional to~$B$ (which is the case for
$\omega<1000\,\mbox{cm}^{-1}$), the magnetic field drops out of
$I_B(\omega_{n+1})$.

\begin{table} \label{SWM} \caption{SWM parameters used for
the calculations}
\begin{ruledtabular}
\begin{tabular}{|c|c|c|}
SWM parameter & full model (eV) & parabolic model (eV) \\
$\gamma_0$ & 3.08  & 3.15\\
$\gamma_1$ & 0.38  & 0.38 ($=\frac{\gamma_1^*}{2}$) \\
$\gamma_3$ & 0.315 & 0 \\
$\gamma_4$ & 0.044 & 0 \\
$\Delta+2\gamma_5-2\gamma_2$ & 0.22  & 0\\
\end{tabular}
\end{ruledtabular}
\end{table}

\begin{figure}
\includegraphics[width=1\linewidth,angle=0,clip]{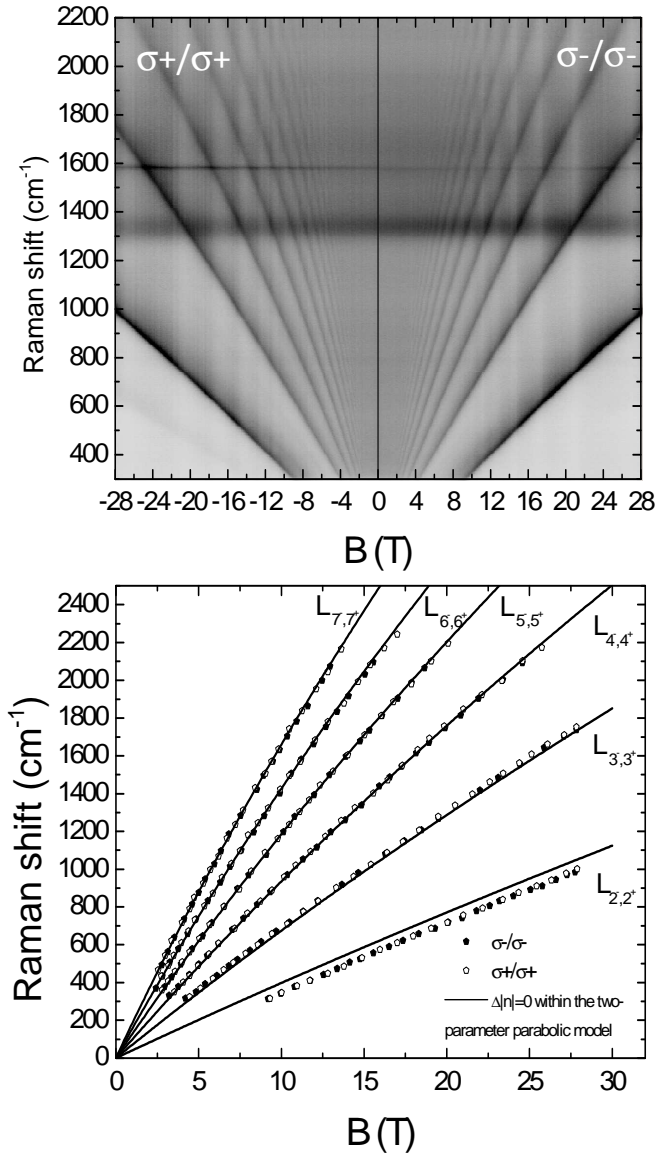}
\caption{\label{FigCoMap} Upper panel: False color map of the
scattered intensity at $T=4.2$~K as a function of the magnetic
field in the two co-circular configurations (raw data). Lower
panel: Position of the maxima of scattered light as a function of
the magnetic field in the $\sigma^+/\sigma^+$ (black dots) and
$\sigma^-/\sigma^-$ (open dots) configurations. Solid black lines
are the $\Delta |n|=0$ electronic excitations calculated within
the two-parameter parabolic model using the parameters described in
Table.~I.}
\end{figure}

The upper panel of Fig.~\ref{FigCoMap} shows a false color map of
the scattered intensity as a function of the magnetic field for
the two $\sigma^-/\sigma^-$ and $\sigma^+/\sigma^+$
configurations. The evolution of these features is nearly linear
with the magnetic field which is expected for carriers with a
parabolic dispersion. The evolution of the maxima of the scattered
light intensity as a function of the magnetic field is presented
in the lower panel of Fig.~\ref{FigCoMap} for the
$\sigma^-/\sigma^-$ (open dots) and $\sigma^+/\sigma^+$ (black
dots) configurations. We present in this figure the results of the
corresponding excitations calculated within the two-parameter
parabolic model with $\gamma_0=3.15$~eV, $\gamma_1^*=0.760$~eV
(black lines). In a first approximation, this model allows to
describe well the overall behavior of excitations involving Landau
levels with $n>2$. The clear limitations of this model appear at
low energies with the $L_{2^-,2^+}$ excitation which is
overestimated, mainly because of the missing $\gamma_3$ parameter
relevant at low energies.

\subsection{Crossed circular configuration}

\begin{figure}
\includegraphics[width=1\linewidth,angle=0,clip]{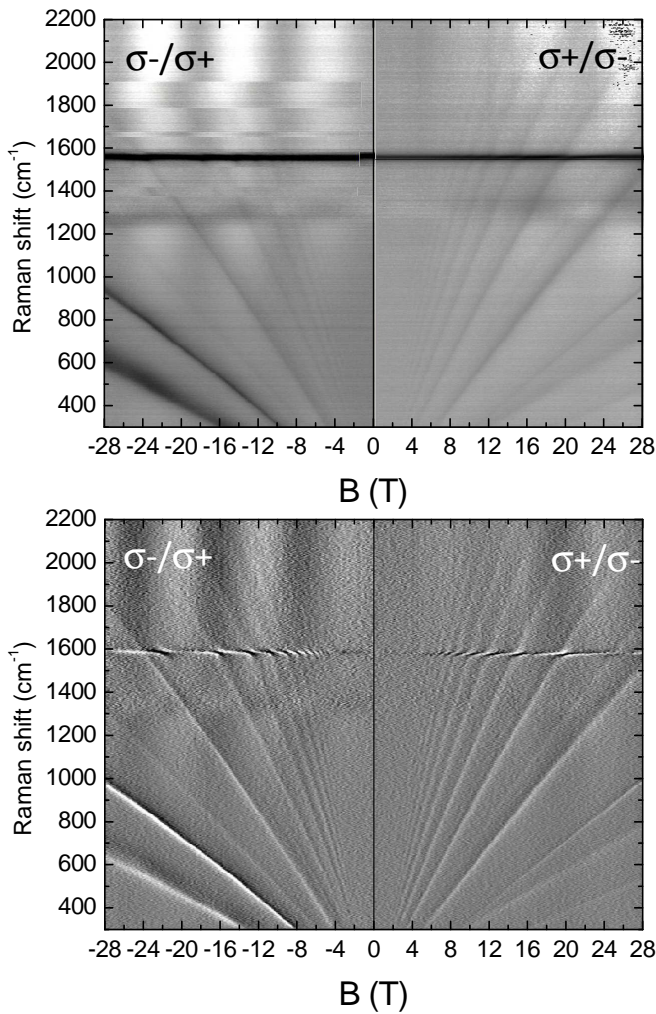}
\caption{\label{FigCrossedMap} Upper panel: False color map of the
scattered intensity at T=4.2K as a function of the magnetic field
in the two crossed circular configurations. Lower panel:
Derivative of the scattered light intensity as a function of the
magnetic field.}
\end{figure}

\begin{figure*}
\includegraphics[width=0.75\linewidth,angle=0,clip]{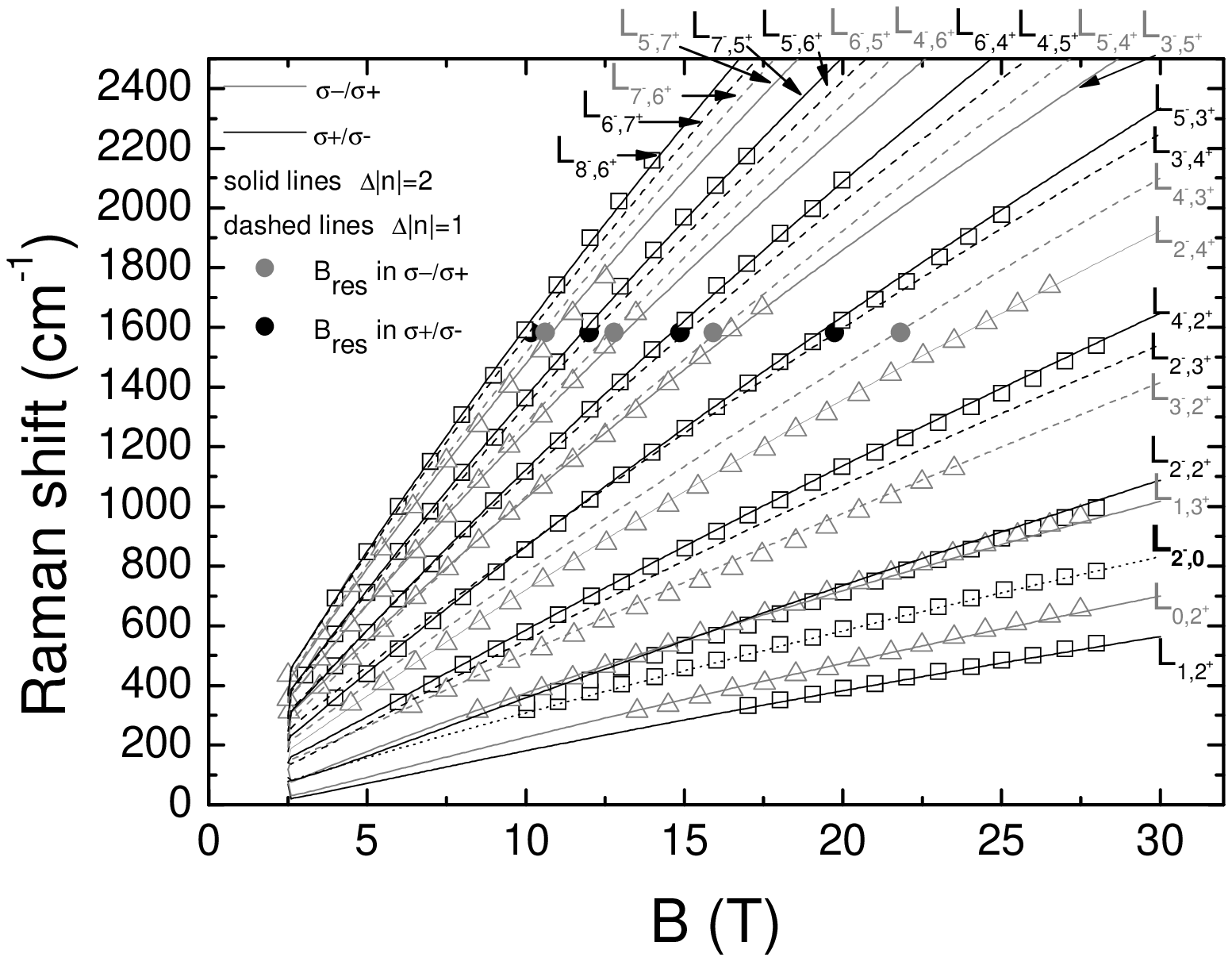}
\caption{\label{DataCrossed} Maxima of scattered intensity as a
function of the magnetic field for $\sigma^-/\sigma^+$ (gray open
triangles) and $\sigma^+/\sigma^-$ (black open squares), together
with $\Delta |n|=\pm 2$ electronic excitations calculated using
the parameters presented in Table.~I within the two-parameter parabolic model
(black and gray solid lines) and $\Delta |n|=\pm 1$ electronic
excitations (black and gray dashed lines). The gray and black
circles correspond to the phonon resonances observed in
$\sigma^+/\sigma^-$ and $\sigma^-/\sigma^+$ configurations
respectively. The line \textbf{$L_{2^-,0}$} is calculated at
$k_z$=0.3.}
\end{figure*}

The crossed circular polarization configuration is expected to
select excitations of the $E_{2g}$ symmetry which includes the
optical phonon ($G$~band feature) and electronic excitations
involving hole and electron states with $\Delta |n|=\pm 2$
(Ref.~\onlinecite{Kashuba2009}), which correspond to angular
momentum $m_z=\pm 2$. We present in the upper panel of
Fig.~\ref{FigCrossedMap} a false-color map of the scattered
intensity as a function of the magnetic field for both crossed
circular polarization configuration and, in the lower panel of the
same figure, a false color map of the derivative with respect to
the magnetic field ($\Delta B = 0.32$~T and spectra acquired every
0.08~T) of the same data. In these figures, many magnetic field
dependent features can be observed, as well as a pronounced
oscillatory behavior of the G-band feature that will be discussed
in detail in the next section. All of these features have an
almost linear dependence on the magnetic field confirming that
$K$-point carriers are involved.

Most of the excitations shown in Fig.~\ref{FigCrossedMap} are
expected to follow the $\Delta |n|=\pm 2$ selection rule. What is
surprising, especially in the $\sigma^+/\sigma^-$ configuration,
is the seeming coincidence of these excitations with the $\Delta
|n|=\pm 1$ excitations which we attribute to the effect of the
trigonal warping discussed in Sec.~\ref{sec:selectionrules} and
which couple to the $E_{2g}$ phonon. Indeed, a closer inspection
of Fig.~\ref{FigCrossedMap} shows that the magnetic field
dependent features that are observed are \emph{not} those that
couple to the phonon. This is particularly visible in the
$\sigma^-/\sigma^+$ configuration where the electron-phonon
interaction is resonant at $B=21.8$~T while the electronic
excitation crosses the phonon energy at $B=24.1$~T. The
coincidence of the magnetic field values corresponding to resonant
electron-phonon interaction revealed by the oscillations of the
phonon and the magnetic field values at which the visible
electronic excitations cross the phonon energy is a result of the
electron-hole asymmetry. This asymmetry leads to nearly
degenerated $(n+1)^-\rightarrow(n-1)^+$ Raman active excitations
and $(n-1)^-\rightarrow n^+$ optical-like excitations. This effect
is less pronounced for lower values of the Landau level index
which makes this particular resonance at high fields well
separated from the Raman active electronic excitation.

In the following, we search for appropriate parameters of the SWM
model, taking into account that (i) $\Delta |n|=\pm 1$ electronic
excitations should cross the phonon energy at the anti-crossing
points and (ii) that the observed electronic excitation should
correspond to the $\Delta |n|=\pm 2$ electronic excitations in
both crossed circular polarization configurations. These two
requirements are indeed fulfilled by the parameters summarized in
Table~I. The results for different electronic excitations
determined from the effective SWM model together with the observed
excitations in both crossed circular polarization configurations
are presented in Fig.~\ref{DataCrossed}. In this figure, the gray
and black lines correspond to the calculated excitations in the
$\sigma^-/\sigma^+$ and $\sigma^+/\sigma^-$ configurations
respectively. The open triangles and open squares correspond to
the maxima of scattered intensity observed in $\sigma^-/\sigma^+$
and $\sigma^+/\sigma^-$ configurations respectively. The gray and
black circles correspond to resonant magnetic fields for the
magneto-phonon effect in these two different configurations. As
can be seen in Fig.~\ref{DataCrossed}, the set of SWM parameters
presented in Table~I describes well most of the observed
excitations both in the co-circular configuration for which the
electron-hole asymmetry does not influence the different
excitations energy and in the crossed circular configuration where
a strong electron-hole asymmetry is observed.

To determine these parameters, we start from the $\gamma_0$ and
$\gamma_1$ values determined from magneto-transmission experiments
on similar samples~\cite{Orlita2009} and from the $\gamma_3$ and
$\gamma_4$ values proposed in Ref.~\onlinecite{Brandt1988}. We
then introduce the electron-hole asymmetry by increasing
$\Delta_{eff}$ starting from values presented in
Ref.~\onlinecite{Brandt1988} and compare the results with
experimental data. The value of $\Delta_{eff}$ is then gradually
increased in order to describe the observed energy difference in
both crossed circular configurations. Introducing the
electron-hole asymmetry has an effect on the $\gamma_0$ value
which have to be slightly decreased (by $\sim 4\%$) to describe
both crossed and co-circular polarization experiments while
$\gamma_1$, $\gamma_3$ and $\gamma_4$ are kept constant. It should
be noted that, even though we have reduced the number of
parameters by introducing $\Delta_{eff}$, there is not a unique
pair of parameters $\gamma_4$ and $\Delta_{eff}$ able to describe
our data as both these parameters affect the asymmetry. The one we
present in Table~I keeps $\gamma_0$ within the accepted values and
allows comparison with other works. For the SWM parameters
proposed by Brandt~\textit{et al.}~\cite{Brandt1988},
$\Delta_{eff}=0.108$~eV, which is, for the same $\gamma_4$ value,
twice lower than the $\Delta_{eff}$ needed to reproduce
experimental results in the present work.

The parameters presented in Table~I are sufficient to describe
$K$-point carriers but are not valid for $k_z \neq 0$ where the
seven SWM parameters need to be specified. One possible
combination is to set $\Delta$ and $\gamma_2$ to their standard
values $\Delta=-0. 008$~eV and
$\gamma_2=-0.02$~eV~\cite{Brandt1988}. This leads to
$\gamma_5=0.094$~eV and of course, at the $K$-point, this set of
parameters brings exactly the same excitation spectrum as the one
calculated using $\Delta_{eff}=0.22$. These parameters can be used
to calculate Landau bands energies at $k_z \neq 0$.

\begin{figure}
\includegraphics[width=0.9\linewidth,angle=0,clip]{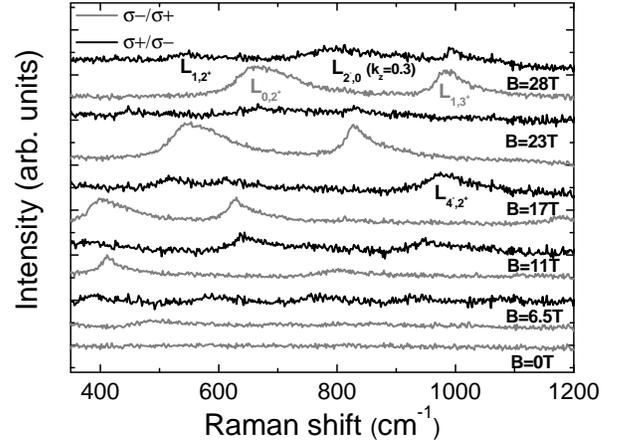}
\caption{\label{FigCrossedSpectre} a) Representative Raman
scattering spectra at different values of the magnetic field in
$\sigma^-/\sigma^+$ (black curves) and $\sigma^+/\sigma^-$ (gray
curves) configurations showing different magnetic field dependent
features.}
\end{figure}

Most of the observed excitation fulfill the Raman scattering
selection rules presented in the previous sections, but we also
observe traces of optical-like excitations with $\Delta |n|=\pm 1$:
$L_{1,2^+}$ in the $\sigma^-/\sigma^+$ configuration and
$L_{3^-,2^+}$ in the $\sigma^+/\sigma^-$.

All electronic excitations observed in the crossed
polarization configuration are significantly weaker than the
symmetric lines observed in co-circular configuration, but
unexpectedly, in the $\sigma^-/\sigma^+$ polarization
configuration two particular features involving the $n=0,1$ Landau
levels are of much stronger intensity. These two features are
presented in Fig.~\ref{FigCrossedSpectre} (gray curves). They have
a significantly different line shape, $L_{1,3^+}$ being
asymmetric as the $\Delta|n|=0$ excitation discussed in the
previous section, while the $L_{0,2^+}$ feature is much broader.
Probably, the difference between these two features and the rest
of $(n-1)^-\to(n+1)^+$ transitions can be explained by the
different $k_z$ dispersion of the $n=0,1$ Landau levels.

In the $\sigma^+/\sigma^-$ polarization configuration $\Delta
|n|=-2$ and possibly $\Delta |n|=+1$ excitations are expected. The
feature with the lowest energy in this configuration is attributed
to $L_{1,2^+}$ excitation at $k_z=0$. The first feature of the
$\Delta |n|=-2$ series is the $L_{2^-,0}$ excitation which should
be blocked at $k_z=0$ due to the full occupation of the $n=0$
level, but $k_z\neq 0$ excitations can be observed. We attribute
the broad feature with a much reduced amplitude as compared to the
others (see Fig~\ref{FigCrossedSpectre}) to the $L_{2^-,0}$
excitation at $k_z=k_F \sim 0.3$ (dotted lines in
Fig.~\ref{DataCrossed}). The second excitation with the same
change of Landau band modulus that could be expected is
$L_{3^-,1}$, which should also be blocked because of occupation
effects in the $n=1$ level. This feature is not clearly observed
in the present experiment.

\begin{figure}
\includegraphics[width=0.7\linewidth,angle=0,clip]{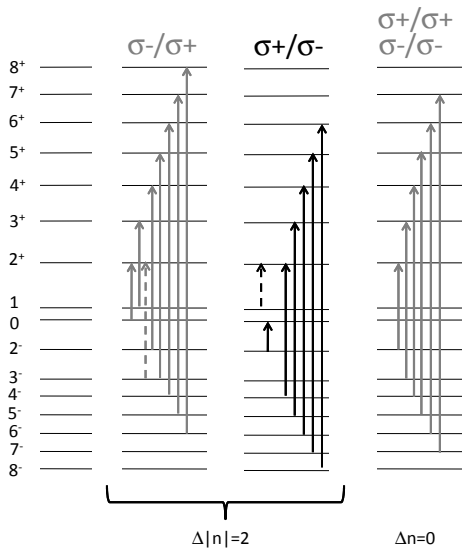}
\caption{\label{SR} Schematics of the observed electronic
excitations in the two crossed circular configurations and in
co-circular configuration in natural graphite. Observed
optical-like excitations are indicated by dashed arrows.}
\end{figure}

To summarize all the obtained results, we present in Fig.~\ref{SR}
a schematic picture of all the observed excitations in the
different crossed- and co-circular polarization configurations. In
the $\sigma^-/\sigma^+$ configuration electronic excitations with
$\Delta|n|=+2$ are observed and traces of the optical-excitation
$L_{3^-,2^+}$. The two excitations involving the $0,1$ Landau
levels also fulfill this index rule. In the $\sigma^+/\sigma^-$,
$\Delta|n|=-2$ excitations are observed together with the
optical-like excitation $L_{1,2^+}$ and of the symmetric
excitation $L_{2^-,2^+}$. The $\Delta|n|=\pm 1$ optical-like
electronic excitations series are not clearly observed in this
configuration but the coupling of these excitations with the
$E_{2g}$ phonon has a strong influence on the phonon feature.

\subsection{Magneto-phonon effect in bulk graphite}

To our knowledge, the magneto-phonon effect involving $K$-point
massive carriers in bulk graphite and the $E_{2g}$ phonon has
never been explored theoretically. This situation is of particular
interest because of the nature of the electronic states involved
in the coupling. The magneto-phonon effect in
graphene~\cite{Ando07, Goerbig2007, Faugeras2009} or in bilayer
graphene~\cite{Ando2007b} involves discrete electronic states
which results in a fully resonant coupling with the phonon. In the
case of $K$-point carriers in graphite, because of the 3D nature
of bulk graphite and of the associated dispersion along $k_z$, the
electronic states belong to Landau bands and the electronic
excitation is spread over a wide range of energy (line width $\sim
50\,\mbox{cm}^{-1}$ in the case of symmetric lines in co-circular
configuration) and the interaction is also spread on such a wide
range of energy. As a result, we expect an effect less pronounced
than in graphene and with asymmetric evolution of the parameters
describing the phonon feature.

\begin{figure}
\includegraphics[width=0.8\linewidth,angle=0,clip]{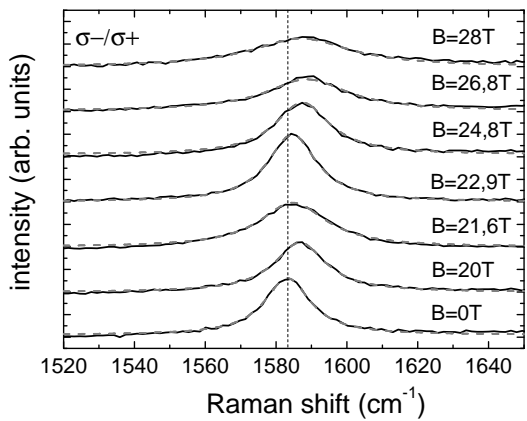}
\caption{\label{SpecPhonon} Raman scattering spectra at selected
values of the magnetic field in the $\sigma^-/\sigma^+$
polarization configuration (solid black curves) and lorentizan
curves (dashed gray curves). Curves are shifted for clarity. The
vertical dashed line indicates the B=0 position.}
\end{figure}

\begin{figure}
\includegraphics[width=0.9\linewidth,angle=0,clip]{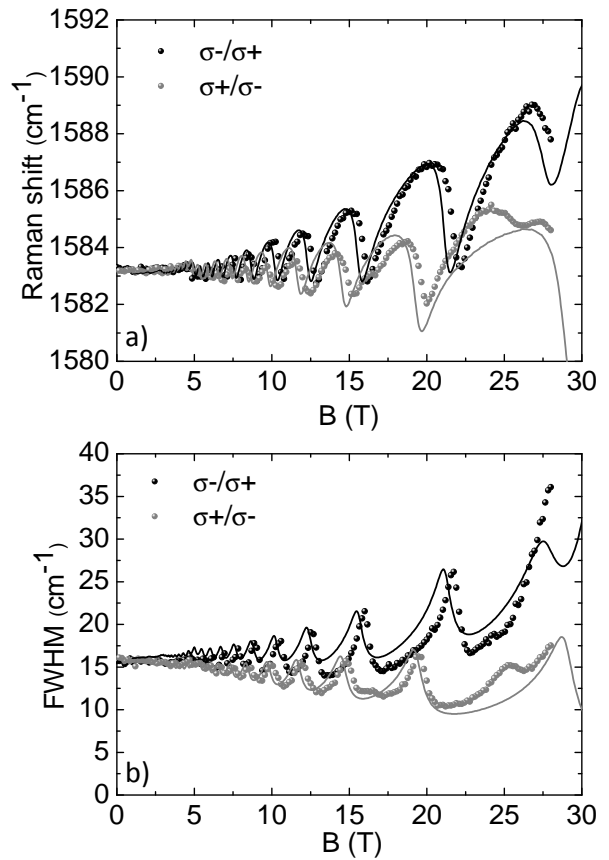}
\caption{\label{Phononmove} a) Raman shift and b) FWHM of the
$E_{2g}$ phonon feature as a function of the magnetic field for
the $\sigma^-/\sigma^+$ (black dots) and $\sigma^+/\sigma^-$ (gray
dots) configurations.}
\end{figure}

As already noticed in the previous section, the $E_{2g}$ phonon
feature shows pronounced oscillations and even an anti-crossing
behavior when optical electronic excitations $\Delta |n|=\pm 1$
(hardly seen in the Raman scattering experiment) are tuned in
resonance with the phonon energy. This can be seen clearly in the
lower panel of Fig.~\ref{FigCrossedMap}. We present in
Fig.~\ref{SpecPhonon} raw Raman scattering spectra in the optical
phonon range of energy for selected values of the magnetic field.
The phonon feature energy and FWHM evolves a lot with magnetic
field and at $B=28$~T a blue shift is clearly visible together
with a strong broadening of the feature. It is possible to use
Lorentzian curves to described these spectra, as shown by the
dashed gray curves in Fig.~\ref{SpecPhonon}. The parameters
(namely, the peak position and its full width at half maximum)
deduced from a Lorentzian fitting of the phonon feature as a
function of the magnetic field are presented in
Fig.~\ref{Phononmove}(a) and (b) for the two crossed polarized
configurations. The oscillations can be observed for magnetic
fields higher than $B\sim 5$~T. One can note on
Fig.~\ref{Phononmove}(a) and (b) that the minima in FWHM or in
Raman shift of the phonon feature in the two opposite crossed
circular polarizations do not occur at the same values of the
magnetic field. This difference is due to the electron-hole
asymmetry in graphite discussed and determined in the previous
section.

To describe the magneto-phonon effect theoretically, we
adopt the same approach that was successfully used for the
magneto-phonon effect in the monolayer
graphene~\cite{Faugeras2009,Yan2010}. Namely, the shift and the
broadening of the phonon are given by the real part and twice the
imaginary part of the complex root~$\omega$ of the equation
\begin{equation}
\omega^2-\omega_0^2=2\omega_0\lambda\left[\Pi_B(\omega+i\Gamma)
-\Pi_{B=0}(\omega_0+i\Gamma)\right],\label{omega=Pi}
\end{equation}
where $\Pi_B(\omega)$ is the polarization operator, which depends
on the magnetic field and frequency. We have subtracted its value
at $B=0$, so the shift is measured with respect to~$\omega_0$, the
phonon frequency at zero field (about $1583\:\mbox{cm}^{-1}$).

The simplest two-parameter parabolic model, used in
Sec.~\ref{cocircular} to describe Raman scattering due to
$\Delta|n|=0$ electronic excitations, is not sufficient to
describe the magneto-phonon effect. Four-band approximation is
needed, and transitions between levels involving the two split-off
bands also need to be taken into account. The only approximation
which is made here is to neglect the trigonal warping by setting
$\gamma_3=0$. For each value of~$B$, $k_z$, and of the Landau
level index~$n$, it is then sufficient to diagonalize a
$4\times{4}$ matrix, instead of an infinite-dimensional one, which
would be necessary for $\gamma_3 \neq 0$. This significantly
improves the computation efficiency, but introduces a certain
error, that will be discussed in the following. Relegating the
explicit expressions to Appendix~A, we plot $\Re\omega$ and
$2\Im\omega$ for the solutions of Eq.~(\ref{omega=Pi}) by solid
lines in Fig.~\ref{Phononmove}. We have adjusted
$\lambda=3.2\times{10}^{-3}$ and $\Gamma=44\:\mbox{cm}^{-1}$ to
reproduce the amplitude of the oscillations and the degree of
asymmetry. We added a constant broadening of $5\:\mbox{cm}^{-1}$
to $2\Im\omega$ which corresponds to a broadening of the phonon by
scattering mechanisms other than the electron-phonon interaction.
We have used the SWM parameters listed in the left column of
Table~I, and set $\gamma_3=0$.

Setting $\gamma_3=0$ changes the inter-Landau-level excitation
frequencies, so their resonance with the phonon occurs at slightly
different values of the magnetic field and this explains why the
minima and maxima of the theoretical and experimental curves are
horizontally shifted in Fig.~\ref{Phononmove}. In principle, one
could remove this discrepancy by slightly readjusting other
SWM parameters. However, this would introduce a correction to the
transition matrix elements, and it is not clear whether they would
be closer to their true values.
Thus, we prefer to stick to the parameters from
Table~I, and consider the discrepancy between the resonance
positions as an order-of-magnitude estimate of the error in the
theory.

The dependence of the magneto-phonon effect on the electronic
Fermi energy,~$\epsilon_F$, deserves a separate discussion. It
enters $\Pi_B(\omega)$ through the filling of Landau levels
$n=0,1$. Namely, all $n^+$ levels with $n\geq{2}$, as well as all
levels from the upper split-off band are assumed to be empty, and
all $n^-$ levels with $n\geq{2}$, as well as all levels from the
lower split-off band are assumed to be filled. The $n=0$ and $n=1$
Landau levels are filled for $k_z<k_{0,1}^F$ and empty at
$k_z>k_{0,1}^F$, where $k_0^F$, $k_1^F$ are those $k_z$ at which
$n=0$ and $n=1$ Landau levels cross the Fermi level,
$\epsilon_0=\epsilon_F$ and $\epsilon_1=\epsilon_F$, respectively.
It turns out that these population effects are responsible for the
smooth in~$B$ component of the curves in Fig.~9, which is
different for the two polarizations (the $\sigma^-/\sigma^+$ curve
exhibits an overall increase in energy with~$B$, while the
$\sigma^+/\sigma^-$ an overall decrease). Indeed, (i)~since there
are no two $1^+$ and $1^-$ Landau levels, but a single one, only
one of the two transitions  $2^-\to{1}$, $1\to{2}^+$ is allowed
for a given~$k_z$, and the other one is blocked by the Pauli
principle (assuming zero temperature); (ii)~as the transitions
$2^-\to{1}$, $1\to{2}^+$ contribute mainly in the
$\sigma^-/\sigma^+$ and $\sigma^-/\sigma^+$ polarization,
respectively, then, depending on the filling of the $n=1$ Landau
level, the contribution is made in only one of the two
polarizations. It turns out that the monotonous in~$B$ component
is mainly due to transitions around $k_z=\pi/2$ ($H$~point), which
becomes resonant with the phonon at fields near $30\,\mbox{T}$.
Indeed, the theoretical curves in Fig.~\ref{Phononmove} are very
little sensitive to the value of $\epsilon_F$ as it is varied from
$2\gamma_2$, the bottom of the conduction band at $k_z=0$ (when
the $n=0,1$ levels are empty for all~$k_z$), to
$\gamma_2+30\,\mbox{meV}$, but when we fully populate the $n=0,1$
levels at all~$k_z$ by pushing $k_{0,1}^F\to\pi/2$, the tendency
is reversed: the $\sigma^-/\sigma^+$ curve decreases in energy
with~$B$, while the $\sigma^+/\sigma^-$ curve increases.

\subsection{Room Temperature Experiments}

\begin{figure}
\includegraphics[width=0.7\linewidth,angle=0,clip]{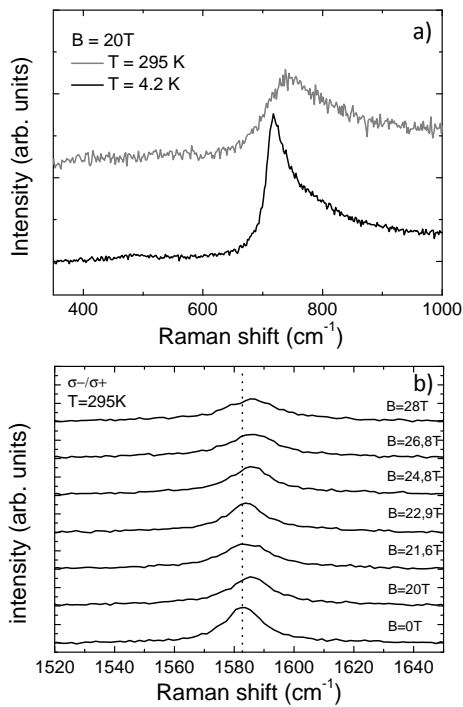}
\caption{\label{RoomTemp} a)  Raman scattering spectra measured at
$T=4.2$~K (black curve) and at $T=295$~K (gray curve) at $B=20$~T
in the $\sigma^+/\sigma^+$ configuration. b) Raman scattering
spectra at selected values of the magnetic field in the
$\sigma^-/\sigma^+$ polarization configuration measured at T=295K.
Curves are shifted for clarity. The vertical dashed line indicates
the B=0 position.}
\end{figure}

The results presented above have been extracted from low
temperature ($4.2$~K) measurements. It is however instructive to
note that that the magneto-Raman scattering response of graphite,
which in fact reflects the quantum character of its energy
structure, survives even at room temperature. We illustrate this
here, although leave a more detailed discussion of temperature
effects to a separate report.

Fig.~\ref{RoomTemp}a) shows the comparison of Raman scattering
response due to $L_{2^-,2^+}$ transition, measured at $4.2$~K and
$295$~K, in co-circular configuration. At room temperature, the
$L_{2^-,2^+}$ spectrum is weaker but still clearly visible. The
shift of the peak towards higher energies and its smearing
observed at room temperature is due to thermal population of the
final state $n=2^+$ Landau band. The low energy onset of the
$L_{2^-,2^+}$ spectrum remains rather sharp at low temperatures
since all the $n=2^+$ Landau band is empty, down to $k_z=0$
states. Thermal population of $n=2^+$ band leads to Pauli blocking
of the transitions, starting from those involving the $k_z=0$
final states, what quantitatively accounts for the shift and
additional broadening of the $L_{2^-,2^+}$ spectrum at elevated
temperatures.

Remarkably, also the magneto-phonon resonance effect in graphite
survives up to the room temperature. Room-temperature Raman
scattering spectra measured in cross-polarization configuration
and focused on the $E_{2g}$ phonon response are shown in
Fig.~\ref{RoomTemp}b). The magneto-oscillations of the $E_{2g}$
phonon are clear at room temperature and resemble very much those
measured at $4.2$~K, presented in Fig.~\ref{SpecPhonon}.

Rough comparison of low and room-temperature data points towards
thermal population effects being the main source of the observed
difference in the spectral response. This implies that other
possible sources of spectral broadening in magneto-Raman
scattering of graphite have a negligible temperature dependence,
extraordinarily, up to the room temperature.

\section{Conclusion}

To conclude, we have used circularly polarized magneto-Raman
scattering techniques in order to study purely electronic and
phonon excitations in bulk graphite. Different series of
electronic excitations arising from $K$-point carriers are
observed. They are interpreted using the Raman scattering
selection rules otherwise characteristic of the bilayer graphene.
The use of polarized Raman scattering reveals a surprisingly
strong electron-hole asymmetry and suggests that the generally
accepted set of SWM parameters from Brandt \textit{et
al.}~\cite{Brandt1988} under estimates the electron-hole
asymmetry. The magneto-phonon effect is observed in the crossed
circularly polarized configuration with different amplitude of the
effect in the two distinct crossed polarized configurations due to
the contribution of the holes at the H point of the graphite
Brillouin zone. Electronic contributions to the Raman scattering
spectrum of bulk graphite and the magneto-phonon effect are
observed up to room temperature.

\section{Acknowledgements}
We acknowledge technical support from Ivan Breslavetz. Part of
this work has been supported by ANR-08-JCJC-0034-01, GACR
P204/10/1020, GRA/10/E006 (EPIGRAT), RTRA "DISPOGRAPH" projects
and by EuroMagNET II under the EU contract number 228043. P.K. is
financially supported by the EU under FP7, contract no. 221515
''MOCNA''. Yu. L. is supported by the Russian state contract No.
16.740.11.0146.

\begin{widetext}

\appendix
\section{Landau levels in the four-band model}

The Hamiltonian of the $AB$-stacked graphite can be written as
\begin{equation}
H= \left(\begin{array}{cccccccccc}
\ldots & \ldots & \ldots & \ldots & \ldots & \ldots & \ldots & \ldots
 & \ldots & \ldots \\
\ldots & V_{11} & V_{12}^\dagger & H_1 & V_{12}^\dagger
& V_{11}^\dagger & 0 & 0 & 0 & \ldots \\
\ldots & 0 & V_{22} & V_{12} & H_2 & V_{12} & V_{22}^\dagger & 0 & 0 &\ldots \\
\ldots & 0 & 0 & V_{11} & V_{12}^\dagger & H_1 & V_{12}^\dagger &
V_{11}^\dagger & 0 &\ldots \\
\ldots & 0 & 0 & 0 & V_{22} & V_{12} & H_2 & V_{12} & V_{22}^\dagger
 & \ldots \\
\ldots & \ldots & \ldots & \ldots & \ldots & \ldots & \ldots &
\ldots & \ldots & \ldots
\end{array}\right),
\end{equation}
where each element is a matrix in the Hilbert space of a single
layer. $H_1$ and $H_2$ contain the in-plane nearest-neighbor
coupling with the matrix element~$\gamma_0$ as well as the
diagonal shift~$\Delta$. $V_{12}$ represents the coupling between
neighboring layers with matrix elements $\gamma_1$ for the
neighboring atoms, and $\gamma_3,\gamma_4$ for the second nearest
neighbors. $V_{11}$ and $V_{22}$ correspond to
second-nearest-layer coupling with matrix elements $\gamma_2/2$
and $\gamma_5/2$. Due to the translational invariance in the $z$
direction, the problem can be reduced to that of an effective
bilayer:
\begin{equation}
H_{k_z}= \left[\begin{array}{cc}
H_1+V_{11}e^{2ik_z}+V_{11}^\dagger{e}^{-2ik_z} &
2V_{12}^\dagger\cos{k}_z \\
2V_{12}\cos{k}_z &
H_2+V_{22}e^{2ik_za_z}+V_{22}^\dagger{e}^{-2ik_za_z}
\end{array}\right].
\end{equation}
Here $k_z$ is the wave vector in the direction perpendicular to
the layers, measured in the units of $1/a_z$, where $a_z$ is the
distance between the neighboring layers. The period of the
structure is $2a_z$, that is, two layers, so
$-\pi/2<{k}_z\leq\pi/2$. Expansion of the single-layer Hamiltonian
in $p_x,p_y$, the in-plane quasi-momentum components counted from
the $H-K-H$ line, gives
\begin{equation}\label{bilayerHam=}
{H}_{k_z}(\hat{\vec{p}})=\left[\begin{array}{cccc} \Gamma_2 &
v\hat{p}_- & -\alpha_4v\hat{p}_- &
\alpha_3v\hat{p}_+ \\
v\hat{p}_+ & \Gamma_5 & \Gamma_1 &
-\alpha_4v\hat{p}_- \\
-\alpha_4v\hat{p}_+ &
\Gamma_1 & \Gamma_5 & v\hat{p}_-\\
\alpha_3v\hat{p}_- & -\alpha_4v\hat{p}_+ & v\hat{p}_+ & \Gamma_2
\end{array}\right],
\end{equation}
where
\begin{eqnarray}
&&v=\frac{3}{2}\,\frac{\gamma_0{a}}{\hbar},\quad
\Gamma_1=2\gamma_1\mathcal{C},\quad
\Gamma_2=2\gamma_2\mathcal{C}^2,\quad
\alpha_{3,4}=\frac{2\gamma_{3,4}}{\gamma_0}\,\mathcal{C},\quad
\Gamma_5=2\gamma_5\mathcal{C}^2+\Delta,\quad
\mathcal{C}\equiv\cos{k}_z,
\label{bilayerparams=}
\end{eqnarray}
$a=1.42\:\mbox{\AA}$ is the distance between the neighboring
carbon atoms in the same layer, and
$\hat{p}_\pm=-i\hbar(\partial_x\pm{i}\partial_y)$.

It is convenient to rotate the basis in the space of the 4-columns
$(\psi_1,\psi_2,\psi_3,\psi_4)^T$ as
\begin{equation}
\psi_1'=\frac{\psi_2+\psi_3}{\sqrt{2}},\quad
\psi_2'=\psi_1,\quad\psi_3'=\psi_4,\quad
\psi_4'=\frac{\psi_2-\psi_3}{\sqrt{2}},
\end{equation}
so the transformed Hamiltonian becomes
\begin{equation}
\label{A5} H'_{k_z}(\hat{\vec{p}})= \left[\begin{array}{cccc}
\Gamma_5+\Gamma_1 & \bar{v}_4\hat{p}_+ &
\bar{v}_4\hat{p}_- & 0 \\
\bar{v}_4\hat{p}_- & \Gamma_2 & \alpha_3v\hat{p}_+ &
v_4\hat{p}_- \\
\bar{v}_4\hat{p}_+ &
\alpha_3v\hat{p}_- & \Gamma_2 & -v_4\hat{p}_+\\
0 & v_4\hat{p}_+ & -v_4\hat{p}_- & \Gamma_5-\Gamma_1
\end{array}\right],
\end{equation}
where $v_4=v(1+\alpha_4)/\sqrt{2}$,
$\bar{v}_4=v(1-\alpha_4)/\sqrt{2}$.

In the presence of a magnetic field, described by the vector
potential in the Landau gauge $A_x=-By$, $A_y=A_z=0$, the wave
functions can be sought in the form
\begin{equation}
\psi_j'(x,y)=e^{ip_xx}\sum_{n=0}^\infty
C_j^{(n)}\Phi_n(y+p_xl_B^2),
\end{equation}
where $\Phi_n(y)$ are the harmonic oscillator wave functions. When
$\gamma_3$ is finite, the coefficient $C_j^{(n)}$ is coupled to
$C_j^{(n+3)}$, which, in turn, is coupled to $C_j^{(n+6)}$, and so
on. In this situation, one can proceed either by perturbation
theory or numerically. If we neglect $\gamma_3$ (set it to zero),
the infinite-dimensional Hamiltonian splits into $4\times{4}$
blocks, corresponding to subspaces defined by
\begin{equation}
\left[\begin{array}{c} \psi_1' \\ \psi_2' \\ \psi_3' \\ \psi_4'
\end{array}\right]
=\left[\begin{array}{c} x_1^{(n)}\Phi_{n-1} \\ x_2^{(n)}\Phi_n \\
x_3^{(n)}\Phi_{n-1} \\ x_4^{(n)}\Phi_{n-1}
\end{array}\right].
\end{equation}
This greatly simplifies the calculation, as in each subspace we
have a only a $4\times{4}$ eigenvalue problem (we denote
$\epsilon_B=v/l_B$):
\begin{equation}
\left[\begin{array}{cccc} \Gamma_5+\Gamma_1-\epsilon &
\sqrt{n}(1-\alpha_4)\epsilon_B &
\sqrt{n-1}(1-\alpha_4)\epsilon_B & 0 \\
\sqrt{n}(1-\alpha_4)\epsilon_B & \Gamma_2-\epsilon & 0 &
\sqrt{n}(1+\alpha_4)\epsilon_B \\
\sqrt{n-1}(1-\alpha_4)\epsilon_B & 0 & \Gamma_2-\epsilon &
-\sqrt{n-1}(1+\alpha_4)\epsilon_B\\
0 & \sqrt{n}(1+\alpha_4)\epsilon_B &
-\sqrt{n-1}(1+\alpha_4)\epsilon_B & \Gamma_5-\Gamma_1-\epsilon
\end{array}\right]
\left[\begin{array}{c} x_1^{(n)} \\ x_2^{(n)} \\ x_3^{(n)} \\
x_4^{(n)}
 \end{array}\right]=0.
\label{H4x4=}
\end{equation}
For $\epsilon_B\ll\Gamma_1$, out of four eigenvalues, the largest
one is of the order of $\Gamma_1$, and the lowest one is of the
order of $-\Gamma_1$. These two correspond to split-off bands and
are denoted by $\epsilon_{n^{++}}$, $\epsilon_{n^{--}}$,
respectively. The other two, $\epsilon_{n^+}$ and
$\epsilon_{n^-}$, correspond to those found in the two-band
low-energy approximation\cite{McCann2006}. Eq.~(\ref{H4x4=}) is
valid at $n\geq{2}$ only. At $n=1$ there are three states
corresponding to $n=1$ with $x_3=0$ and energies
$\epsilon_{1^{++}}\sim\Gamma_1$ $\epsilon_{1^{--}}\sim-\Gamma_1$,
and $\epsilon_1$ with low energy. There is one state corresponding
to $n=0$, with $x_1=x_3=x_4=0$ and $\epsilon_0=\Gamma_2$.

Optical transitions for each circular polarization are described
by the operators
$\hat{v}_\pm=(\hat{v}_x\pm{i}\hat{v}_y)/\sqrt{2}$, where
$\hat{v}_x,\hat{v}_y$ are the components of the electronic
velocity operator,
$\hat{\vec{v}}=\partial{H}_{k_z}'(\hat{\vec{p}})/\partial\hat{\vec{p}}$.
The operator $\hat{v}_+$ has matrix elements only between the $n$
and $n-1$ manifolds:
\begin{eqnarray}
&&\langle{n}-1|\frac{\hat{v}_+}{v}|n\rangle=
(1-\alpha_4)\left(x_1^{(n-1)}x_3^{(n)}+x_2^{(n-1)}x_1^{(n)}\right)
+(1+\alpha_4)\left(x_2^{(n-1)}x_4^{(n)}-x_4^{(n-1)}x_3^{(n)}\right),
\quad n\geq{2},\\
&&\langle{0}|(\hat{v}_+/v)|1\rangle
=(1-\alpha_4)x_1^{(1)}+(1+\alpha_4)x_4^{(1)}.
\end{eqnarray}
To obtain the polarization operator $\Pi_B(\omega)$, one has to
sum up contributions from all~$k_z$:
\begin{equation}
\Pi_B(\omega)=-\frac{1}{l_B^2}
\int\limits_{-\pi/2}^{\pi/2}\frac{dk_z}\pi\,
\sum_{n=0}^{\infty}\mathcal{P}_n(\omega,k_z),\label{PiB=}
\end{equation}
where each $\mathcal{P}_n$ contains transitions between Landau
levels from the $n$th and $(n+1)$st manifolds:
\begin{eqnarray}
&&\mathcal{P}_{n\geq{2}}
=\frac{|\langle{n}^+|\hat{v}_+|(n+1)^-\rangle|^2}%
{\epsilon_{n^+}-\epsilon_{(n+1)^-}\mp\omega}
+\frac{|\langle{n}^{++}|\hat{v}_+|(n+1)^-\rangle|^2}%
{\epsilon_{n^{++}}-\epsilon_{(n+1)^-}\mp\omega}
+\frac{|\langle{n}^+|\hat{v}_+|(n+1)^{--}\rangle|^2}%
{\epsilon_{n^+}-\epsilon_{(n+1)^{--}}\mp\omega}
+\frac{|\langle{n}^{++}|\hat{v}_+|(n+1)^{--}\rangle|^2}%
{\epsilon_{n^{++}}-\epsilon_{(n+1)^{--}}\mp\omega}+
\nonumber\\
&&\hspace*{1cm}
{}+\frac{|\langle{n}^-|\hat{v}_+|(n+1)^+\rangle|^2}%
{\epsilon_{(n+1)^+}-\epsilon_{n^-}\pm\omega}
+\frac{|\langle{n}^-|\hat{v}_+|(n+1)^{++}\rangle|^2}%
{\epsilon_{(n+1)^{++}}-\epsilon_{n^-}\pm\omega}
+\frac{|\langle{n}^{--}|\hat{v}_+|(n+1)^+\rangle|^2}%
{\epsilon_{(n+1)^+}-\epsilon_{n^{--}}\pm\omega}
+\frac{|\langle{n}^{--}|\hat{v}_+|(n+1)^{++}\rangle|^2}%
{\epsilon_{(n+1)^{++}}-\epsilon_{n^{--}}\pm\omega},\qquad\\
&&\mathcal{P}_1
=\theta(k_z-k_1^F)\left(\frac{|\langle{1}|\hat{v}_+|2^-\rangle|^2}%
{\epsilon_1-\epsilon_{2^-}\mp\omega}
+\frac{|\langle{1}|\hat{v}_+|2^{--}\rangle|^2}%
{\epsilon_1-\epsilon_{2^{--}}\mp\omega}\right)
+\frac{|\langle{1}^{++}|\hat{v}_+|2^{--}\rangle|^2}%
{\epsilon_{1^{++}}-\epsilon_{2^{--}}\mp\omega}+\nonumber\\
&&\hspace*{1cm} {}+\theta(k_1^F-k_z)\left(
\frac{|\langle{1}|\hat{v}_+|2^+\rangle|^2}%
{\epsilon_{2^+}-\epsilon_1\pm\omega}
+\frac{|\langle{1}|\hat{v}_+|2^{++}\rangle|^2}%
{\epsilon_{2^{++}}-\epsilon_1\pm\omega}\right)
+\frac{|\langle{1}^{--}|\hat{v}_+|2^{++}\rangle|^2}%
{\epsilon_{2^{++}}-\epsilon_{1^{--}}\pm\omega},\\
&&\mathcal{P}_0
=\theta(k_z-k_0^F)\,\frac{|\langle{0}|\hat{v}_+|1^{--}\rangle|^2}%
{\epsilon_0-\epsilon_{1^{--}}\mp\omega}
+\theta(k_0^F-k_z)\,\frac{|\langle{0}|\hat{v}_+|1^{++}\rangle|^2}%
{\epsilon_{1^{++}}-\epsilon_0\pm\omega}
+\theta(k_z-k_1^F)\,\theta(k_0^F-k_z)\,
\frac{|\langle{0}|\hat{v}_+|1\rangle|^2}%
{\epsilon_1-\epsilon_0\pm\omega}.
\end{eqnarray}
The upper and lower $\pm$ signs in the denominators correspond to
$\sigma^-/\sigma^+$ and $\sigma^+/\sigma^-$ polarizations,
respectively. Both the electronic energies and the velocity matrix
elements depend on~$B$ and~$k_z$. The Fermi wave vectors
$k_0^F(B)$, $k_1^F(B)$ are those $k_z$ at which $n=0$ and $n=1$
Landau levels match the Fermi energy, $\epsilon_0=\epsilon_F$ and
$\epsilon_1=\epsilon_F$, respectively. The sum over~$n$ in
Eq.~(\ref{PiB=}) is divergent, so we cut it off at a certain
energy~$\epsilon_\mathrm{max}$. Namely, for transitions not
involving the split-off bands, we sum over those~$n$ for which
$\epsilon_{(n+1)^+}<\epsilon_\mathrm{max}$. Transitions involving
the split-off bands are counted if
$\epsilon_{(n+1)^{++}}<\epsilon_\mathrm{max}$. The subsequent
integration over~$k_z$ is performed numerically. The results
presented in Sec.~V were obtained with
$\epsilon_\mathrm{max}=1\,\mbox{eV}$, and we have verified that
the difference $\Pi_B(\omega)-\Pi_{B=0}(\omega_0)$ does not depend
on $\epsilon_\mathrm{max}$.
\end{widetext}


\bigskip

\begin{thebibliography}{39}
\makeatletter
\providecommand \@ifxundefined [1]{%
 \@ifx{#1\undefined}
}%
\providecommand \@ifnum [1]{%
 \ifnum #1\expandafter \@firstoftwo
 \else \expandafter \@secondoftwo
 \fi
}%
\providecommand \@ifx [1]{%
 \ifx #1\expandafter \@firstoftwo
 \else \expandafter \@secondoftwo
 \fi
}%
\providecommand \natexlab [1]{#1}%
\providecommand \enquote  [1]{``#1''}%
\providecommand \bibnamefont  [1]{#1}%
\providecommand \bibfnamefont [1]{#1}%
\providecommand \citenamefont [1]{#1}%
\providecommand \href@noop [0]{\@secondoftwo}%
\providecommand \href [0]{\begingroup \@sanitize@url \@href}%
\providecommand \@href[1]{\@@startlink{#1}\@@href}%
\providecommand \@@href[1]{\endgroup#1\@@endlink}%
\providecommand \@sanitize@url [0]{\catcode `\\12\catcode
`\$12\catcode
  `\&12\catcode `\#12\catcode `\^12\catcode `\_12\catcode `\%12\relax}%
\providecommand \@@startlink[1]{}%
\providecommand \@@endlink[0]{}%
\providecommand \url  [0]{\begingroup\@sanitize@url \@url }%
\providecommand \@url [1]{\endgroup\@href {#1}{\urlprefix }}%
\providecommand \urlprefix  [0]{URL }%
\providecommand \Eprint [0]{\href }%
\providecommand \doibase [0]{http://dx.doi.org/}%
\providecommand \selectlanguage [0]{\@gobble}%
\providecommand \bibinfo  [0]{\@secondoftwo}%
\providecommand \bibfield  [0]{\@secondoftwo}%
\providecommand \translation [1]{[#1]}%
\providecommand \BibitemOpen [0]{}%
\providecommand \bibitemStop [0]{}%
\providecommand \bibitemNoStop [0]{.\EOS\space}%
\providecommand \EOS [0]{\spacefactor3000\relax}%
\providecommand \BibitemShut  [1]{\csname bibitem#1\endcsname}%
\let\auto@bib@innerbib\@empty
\bibitem [{\citenamefont {Galt}\ \emph {et~al.}(1956)\citenamefont {Galt},
  \citenamefont {Yager},\ and\ \citenamefont {Dail}}]{Galt1956}%
  \BibitemOpen
  \bibfield  {author} {\bibinfo {author} {\bibfnamefont {J.}~\bibnamefont
  {Galt}}, \bibinfo {author} {\bibfnamefont {W.}~\bibnamefont {Yager}}, \ and\
  \bibinfo {author} {\bibfnamefont {H.}~\bibnamefont {Dail}},\ }\href@noop {}
  {\bibfield  {journal} {\bibinfo  {journal} {Phys. Rev.}\ }\textbf {\bibinfo
  {volume} {103}},\ \bibinfo {pages} {1586} (\bibinfo {year}
  {1956})}\BibitemShut {NoStop}%
\bibitem [{\citenamefont {Schroeder}\ \emph {et~al.}(1968)\citenamefont
  {Schroeder}, \citenamefont {Dresselhaus},\ and\ \citenamefont
  {Javan}}]{Schroeder1968}%
  \BibitemOpen
  \bibfield  {author} {\bibinfo {author} {\bibfnamefont {P.~R.}\ \bibnamefont
  {Schroeder}}, \bibinfo {author} {\bibfnamefont {M.~S.}\ \bibnamefont
  {Dresselhaus}}, \ and\ \bibinfo {author} {\bibfnamefont {A.}~\bibnamefont
  {Javan}},\ }\href {\doibase 10.1103/PhysRevLett.20.1292} {\bibfield
  {journal} {\bibinfo  {journal} {Phys. Rev. Lett.}\ }\textbf {\bibinfo
  {volume} {20}},\ \bibinfo {pages} {1292} (\bibinfo {year}
  {1968})}\BibitemShut {NoStop}%
\bibitem [{\citenamefont {Toy}\ \emph {et~al.}(1977)\citenamefont {Toy},
  \citenamefont {Dresselhaus},\ and\ \citenamefont {Dresselhaus}}]{Toy1977}%
  \BibitemOpen
  \bibfield  {author} {\bibinfo {author} {\bibfnamefont {W.~W.}\ \bibnamefont
  {Toy}}, \bibinfo {author} {\bibfnamefont {M.~S.}\ \bibnamefont
  {Dresselhaus}}, \ and\ \bibinfo {author} {\bibfnamefont {G.}~\bibnamefont
  {Dresselhaus}},\ }\href@noop {} {\bibfield  {journal} {\bibinfo  {journal}
  {Phys. Rev. B}\ }\textbf {\bibinfo {volume} {15}},\ \bibinfo {pages} {4077}
  (\bibinfo {year} {1977})}\BibitemShut {NoStop}%
\bibitem [{\citenamefont {Li}\ \emph {et~al.}(2006)\citenamefont {Li},
  \citenamefont {Tsai}, \citenamefont {Padilla}, \citenamefont {Dordevic},
  \citenamefont {Burch}, \citenamefont {Wang},\ and\ \citenamefont
  {Basov}}]{Li2006}%
  \BibitemOpen
  \bibfield  {author} {\bibinfo {author} {\bibfnamefont {Z.~Q.}\ \bibnamefont
  {Li}}, \bibinfo {author} {\bibfnamefont {S.-W.}\ \bibnamefont {Tsai}},
  \bibinfo {author} {\bibfnamefont {W.~J.}\ \bibnamefont {Padilla}}, \bibinfo
  {author} {\bibfnamefont {S.~V.}\ \bibnamefont {Dordevic}}, \bibinfo {author}
  {\bibfnamefont {K.~S.}\ \bibnamefont {Burch}}, \bibinfo {author}
  {\bibfnamefont {Y.~J.}\ \bibnamefont {Wang}}, \ and\ \bibinfo {author}
  {\bibfnamefont {D.~N.}\ \bibnamefont {Basov}},\ }\href@noop {} {\bibfield
  {journal} {\bibinfo  {journal} {Phys. Rev. B}\ }\textbf {\bibinfo {volume}
  {74}},\ \bibinfo {pages} {195404} (\bibinfo {year} {2006})}\BibitemShut
  {NoStop}%
\bibitem [{\citenamefont {Sadowski}\ \emph {et~al.}(2006)\citenamefont
  {Sadowski}, \citenamefont {Martinez}, \citenamefont {Potemski}, \citenamefont
  {Berger},\ and\ \citenamefont {de~Heer}}]{Sadowski06}%
  \BibitemOpen
  \bibfield  {author} {\bibinfo {author} {\bibfnamefont {M.~L.}\ \bibnamefont
  {Sadowski}}, \bibinfo {author} {\bibfnamefont {G.}~\bibnamefont {Martinez}},
  \bibinfo {author} {\bibfnamefont {M.}~\bibnamefont {Potemski}}, \bibinfo
  {author} {\bibfnamefont {C.}~\bibnamefont {Berger}}, \ and\ \bibinfo {author}
  {\bibfnamefont {W.~A.}\ \bibnamefont {de~Heer}},\ }\href@noop {} {\bibfield
  {journal} {\bibinfo  {journal} {Phys. Rev. Lett.}\ }\textbf {\bibinfo
  {volume} {97}},\ \bibinfo {pages} {266405} (\bibinfo {year}
  {2006})}\BibitemShut {NoStop}%
\bibitem [{\citenamefont {Jiang}\ \emph {et~al.}(2007)\citenamefont {Jiang},
  \citenamefont {Henriksen}, \citenamefont {Tung}, \citenamefont {Wang},
  \citenamefont {Schwartz}, \citenamefont {Han}, \citenamefont {Kim},\ and\
  \citenamefont {Stormer}}]{Jiang07}%
  \BibitemOpen
  \bibfield  {author} {\bibinfo {author} {\bibfnamefont {Z.}~\bibnamefont
  {Jiang}}, \bibinfo {author} {\bibfnamefont {E.~A.}\ \bibnamefont
  {Henriksen}}, \bibinfo {author} {\bibfnamefont {L.~C.}\ \bibnamefont {Tung}},
  \bibinfo {author} {\bibfnamefont {Y.~J.}\ \bibnamefont {Wang}}, \bibinfo
  {author} {\bibfnamefont {M.~E.}\ \bibnamefont {Schwartz}}, \bibinfo {author}
  {\bibfnamefont {M.~Y.}\ \bibnamefont {Han}}, \bibinfo {author} {\bibfnamefont
  {P.}~\bibnamefont {Kim}}, \ and\ \bibinfo {author} {\bibfnamefont {H.~L.}\
  \bibnamefont {Stormer}},\ }\href@noop {} {\bibfield  {journal} {\bibinfo
  {journal} {Phys. Rev. Lett.}\ }\textbf {\bibinfo {volume} {98}},\ \bibinfo
  {pages} {197403} (\bibinfo {year} {2007})}\BibitemShut {NoStop}%
\bibitem [{\citenamefont {Orlita}\ \emph
  {et~al.}(2008{\natexlab{a}})\citenamefont {Orlita}, \citenamefont {Faugeras},
  \citenamefont {Plochocka}, \citenamefont {Neugebauer}, \citenamefont
  {Martinez}, \citenamefont {Maude}, \citenamefont {Barra}, \citenamefont
  {Sprinkle}, \citenamefont {Berger}, \citenamefont {de~Heer},\ and\
  \citenamefont {Potemski}}]{Orlita08}%
  \BibitemOpen
  \bibfield  {author} {\bibinfo {author} {\bibfnamefont {M.}~\bibnamefont
  {Orlita}}, \bibinfo {author} {\bibfnamefont {C.}~\bibnamefont {Faugeras}},
  \bibinfo {author} {\bibfnamefont {P.}~\bibnamefont {Plochocka}}, \bibinfo
  {author} {\bibfnamefont {P.}~\bibnamefont {Neugebauer}}, \bibinfo {author}
  {\bibfnamefont {G.}~\bibnamefont {Martinez}}, \bibinfo {author}
  {\bibfnamefont {D.~K.}\ \bibnamefont {Maude}}, \bibinfo {author}
  {\bibfnamefont {A.~L.}\ \bibnamefont {Barra}}, \bibinfo {author}
  {\bibfnamefont {M.}~\bibnamefont {Sprinkle}}, \bibinfo {author}
  {\bibfnamefont {C.}~\bibnamefont {Berger}}, \bibinfo {author} {\bibfnamefont
  {W.~A.}\ \bibnamefont {de~Heer}}, \ and\ \bibinfo {author} {\bibfnamefont
  {M.}~\bibnamefont {Potemski}},\ }\href@noop {} {\bibfield  {journal}
  {\bibinfo  {journal} {Phys. Rev. Lett.}\ }\textbf {\bibinfo {volume} {101}},\
  \bibinfo {pages} {267601} (\bibinfo {year} {2008}{\natexlab{a}})}\BibitemShut
  {NoStop}%
\bibitem [{\citenamefont {Orlita}\ \emph
  {et~al.}(2008{\natexlab{b}})\citenamefont {Orlita}, \citenamefont {Faugeras},
  \citenamefont {Martinez}, \citenamefont {Maude}, \citenamefont {Sadowski},\
  and\ \citenamefont {Potemski}}]{Orlita08a}%
  \BibitemOpen
  \bibfield  {author} {\bibinfo {author} {\bibfnamefont {M.}~\bibnamefont
  {Orlita}}, \bibinfo {author} {\bibfnamefont {C.}~\bibnamefont {Faugeras}},
  \bibinfo {author} {\bibfnamefont {G.}~\bibnamefont {Martinez}}, \bibinfo
  {author} {\bibfnamefont {D.~K.}\ \bibnamefont {Maude}}, \bibinfo {author}
  {\bibfnamefont {M.~L.}\ \bibnamefont {Sadowski}}, \ and\ \bibinfo {author}
  {\bibfnamefont {M.}~\bibnamefont {Potemski}},\ }\href@noop {} {\bibfield
  {journal} {\bibinfo  {journal} {Phys. Rev. Lett.}\ }\textbf {\bibinfo
  {volume} {100}},\ \bibinfo {pages} {136403} (\bibinfo {year}
  {2008}{\natexlab{b}})}\BibitemShut {NoStop}%
\bibitem [{\citenamefont {Henriksen}\ \emph {et~al.}(2007)\citenamefont
  {Henriksen}, \citenamefont {Jiang}, \citenamefont {Tung}, \citenamefont
  {Takita}, \citenamefont {Wang}, \citenamefont {Kim},\ and\ \citenamefont
  {Stormer}}]{Henriksen2008}%
  \BibitemOpen
  \bibfield  {author} {\bibinfo {author} {\bibfnamefont {E.~A.}\ \bibnamefont
  {Henriksen}}, \bibinfo {author} {\bibfnamefont {Z.}~\bibnamefont {Jiang}},
  \bibinfo {author} {\bibfnamefont {L.~C.}\ \bibnamefont {Tung}}, \bibinfo
  {author} {\bibfnamefont {M.~E. S.~M.}\ \bibnamefont {Takita}}, \bibinfo
  {author} {\bibfnamefont {Y.~J.}\ \bibnamefont {Wang}}, \bibinfo {author}
  {\bibfnamefont {P.}~\bibnamefont {Kim}}, \ and\ \bibinfo {author}
  {\bibfnamefont {H.~L.}\ \bibnamefont {Stormer}},\ }\href@noop {} {\bibfield
  {journal} {\bibinfo  {journal} {Phys. Rev. Lett.}\ }\textbf {\bibinfo
  {volume} {100}},\ \bibinfo {pages} {087403} (\bibinfo {year}
  {2007})}\BibitemShut {NoStop}%
\bibitem [{\citenamefont {Orlita}\ \emph {et~al.}(2009)\citenamefont {Orlita},
  \citenamefont {Faugeras}, \citenamefont {Schneider}, \citenamefont
  {Martinez}, \citenamefont {Maude},\ and\ \citenamefont
  {Potemski}}]{Orlita2009}%
  \BibitemOpen
  \bibfield  {author} {\bibinfo {author} {\bibfnamefont {M.}~\bibnamefont
  {Orlita}}, \bibinfo {author} {\bibfnamefont {C.}~\bibnamefont {Faugeras}},
  \bibinfo {author} {\bibfnamefont {J.}~\bibnamefont {Schneider}}, \bibinfo
  {author} {\bibfnamefont {G.}~\bibnamefont {Martinez}}, \bibinfo {author}
  {\bibfnamefont {D.~K.}\ \bibnamefont {Maude}}, \ and\ \bibinfo {author}
  {\bibfnamefont {M.}~\bibnamefont {Potemski}},\ }\href@noop {} {\bibfield
  {journal} {\bibinfo  {journal} {Phys. Rev. Lett.}\ }\textbf {\bibinfo
  {volume} {102}},\ \bibinfo {pages} {166401} (\bibinfo {year}
  {2009})}\BibitemShut {NoStop}%
\bibitem [{\citenamefont {Chuang}\ \emph {et~al.}(2009)\citenamefont {Chuang},
  \citenamefont {Baker},\ and\ \citenamefont {Nicholas}}]{Chuang2009}%
  \BibitemOpen
  \bibfield  {author} {\bibinfo {author} {\bibfnamefont {K.~C.}\ \bibnamefont
  {Chuang}}, \bibinfo {author} {\bibfnamefont {A.~M.~R.}\ \bibnamefont
  {Baker}}, \ and\ \bibinfo {author} {\bibfnamefont {R.~J.}\ \bibnamefont
  {Nicholas}},\ }\href@noop {} {\bibfield  {journal} {\bibinfo  {journal}
  {Phys. Rev. B}\ }\textbf {\bibinfo {volume} {80}},\ \bibinfo {pages}
  {161410(R)} (\bibinfo {year} {2009})}\BibitemShut {NoStop}%
\bibitem [{\citenamefont {Orlita}\ and\ \citenamefont
  {Potemski}(2010)}]{Orlita2010}%
  \BibitemOpen
  \bibfield  {author} {\bibinfo {author} {\bibfnamefont {M.}~\bibnamefont
  {Orlita}}\ and\ \bibinfo {author} {\bibfnamefont {M.}~\bibnamefont
  {Potemski}},\ }\href@noop {} {\bibfield  {journal} {\bibinfo  {journal}
  {Semicond. Sci. Technol.}\ }\textbf {\bibinfo {volume} {25}},\ \bibinfo
  {pages} {063001} (\bibinfo {year} {2010})}\BibitemShut {NoStop}%
\bibitem [{\citenamefont {Faugeras}\ \emph {et~al.}(2011)\citenamefont
  {Faugeras}, \citenamefont {Amado}, \citenamefont {Kossacki}, \citenamefont
  {Orlita}, \citenamefont {Kuhne}, \citenamefont {Nicolet}, \citenamefont
  {Latyshev},\ and\ \citenamefont {Potemski}}]{Faugeras2011}%
  \BibitemOpen
  \bibfield  {author} {\bibinfo {author} {\bibfnamefont {C.}~\bibnamefont
  {Faugeras}}, \bibinfo {author} {\bibfnamefont {M.}~\bibnamefont {Amado}},
  \bibinfo {author} {\bibfnamefont {P.}~\bibnamefont {Kossacki}}, \bibinfo
  {author} {\bibfnamefont {M.}~\bibnamefont {Orlita}}, \bibinfo {author}
  {\bibfnamefont {M.}~\bibnamefont {Kuhne}}, \bibinfo {author} {\bibfnamefont
  {A.}~\bibnamefont {Nicolet}}, \bibinfo {author} {\bibfnamefont
  {Y.}~\bibnamefont {Latyshev}}, \ and\ \bibinfo {author} {\bibfnamefont
  {M.}~\bibnamefont {Potemski}},\ }\href@noop {} {\bibfield  {journal}
  {\bibinfo  {journal} {Phys. Rev. Lett.}\ }\textbf {\bibinfo {volume} {107}},\
  \bibinfo {pages} {036807} (\bibinfo {year} {2011})}\BibitemShut {NoStop}%
\bibitem [{\citenamefont {Dresselhaus}\ \emph {et~al.}(2010)\citenamefont
  {Dresselhaus}, \citenamefont {Jorio},\ and\ \citenamefont
  {Saito}}]{Dresselhaus2010}%
  \BibitemOpen
  \bibfield  {author} {\bibinfo {author} {\bibfnamefont {M.}~\bibnamefont
  {Dresselhaus}}, \bibinfo {author} {\bibfnamefont {A.}~\bibnamefont {Jorio}},
  \ and\ \bibinfo {author} {\bibfnamefont {R.}~\bibnamefont {Saito}},\
  }\href@noop {} {\bibfield  {journal} {\bibinfo  {journal} {Annu. Rev.
  Condens. Matter Phys.}\ }\textbf {\bibinfo {volume} {1}},\ \bibinfo {pages}
  {89} (\bibinfo {year} {2010})}\BibitemShut {NoStop}%
\bibitem [{\citenamefont {Ando}(2007{\natexlab{a}})}]{Ando07}%
  \BibitemOpen
  \bibfield  {author} {\bibinfo {author} {\bibfnamefont {T.}~\bibnamefont
  {Ando}},\ }\href@noop {} {\bibfield  {journal} {\bibinfo  {journal} {J. Phys.
  Soc. Jpn.}\ }\textbf {\bibinfo {volume} {76}},\ \bibinfo {pages} {024712}
  (\bibinfo {year} {2007}{\natexlab{a}})}\BibitemShut {NoStop}%
\bibitem [{\citenamefont {Goerbig}\ \emph {et~al.}(2007)\citenamefont
  {Goerbig}, \citenamefont {Fuchs}, \citenamefont {Kechedzhi},\ and\
  \citenamefont {Fal'ko}}]{Goerbig2007}%
  \BibitemOpen
  \bibfield  {author} {\bibinfo {author} {\bibfnamefont {M.~O.}\ \bibnamefont
  {Goerbig}}, \bibinfo {author} {\bibfnamefont {J.~N.}\ \bibnamefont {Fuchs}},
  \bibinfo {author} {\bibfnamefont {K.}~\bibnamefont {Kechedzhi}}, \ and\
  \bibinfo {author} {\bibfnamefont {V.}~\bibnamefont {Fal'ko}},\ }\href@noop {}
  {\bibfield  {journal} {\bibinfo  {journal} {Phys. Rev. Lett.}\ }\textbf
  {\bibinfo {volume} {99}},\ \bibinfo {pages} {087402} (\bibinfo {year}
  {2007})}\BibitemShut {NoStop}%
\bibitem [{\citenamefont {Kashuba}\ and\ \citenamefont
  {Fal'ko}(2009)}]{Kashuba2009}%
  \BibitemOpen
  \bibfield  {author} {\bibinfo {author} {\bibfnamefont {O.}~\bibnamefont
  {Kashuba}}\ and\ \bibinfo {author} {\bibfnamefont {V.~I.}\ \bibnamefont
  {Fal'ko}},\ }\href@noop {} {\bibfield  {journal} {\bibinfo  {journal} {Phys.
  Rev. B}\ }\textbf {\bibinfo {volume} {80}},\ \bibinfo {pages} {241404(R)}
  (\bibinfo {year} {2009})}\BibitemShut {NoStop}%
\bibitem [{\citenamefont {Mucha-Kruczynski}\ \emph {et~al.}(2010)\citenamefont
  {Mucha-Kruczynski}, \citenamefont {Kashuba},\ and\ \citenamefont
  {Fal'ko}}]{Mucha-Kruczynski2010}%
  \BibitemOpen
  \bibfield  {author} {\bibinfo {author} {\bibfnamefont {M.}~\bibnamefont
  {Mucha-Kruczynski}}, \bibinfo {author} {\bibfnamefont {O.}~\bibnamefont
  {Kashuba}}, \ and\ \bibinfo {author} {\bibfnamefont {V.~I.}\ \bibnamefont
  {Fal'ko}},\ }\href@noop {} {\bibfield  {journal} {\bibinfo  {journal} {Phys.
  Rev. B}\ }\textbf {\bibinfo {volume} {82}},\ \bibinfo {pages} {045405}
  (\bibinfo {year} {2010})}\BibitemShut {NoStop}%
\bibitem [{\citenamefont {Faugeras}\ \emph {et~al.}(2009)\citenamefont
  {Faugeras}, \citenamefont {Kossacki}, \citenamefont {Amado}, \citenamefont
  {Orlita}, \citenamefont {Sprinkle}, \citenamefont {Berger}, \citenamefont
  {de~Heer},\ and\ \citenamefont {Potemski}}]{Faugeras2009}%
  \BibitemOpen
  \bibfield  {author} {\bibinfo {author} {\bibfnamefont {C.}~\bibnamefont
  {Faugeras}}, \bibinfo {author} {\bibfnamefont {P.}~\bibnamefont {Kossacki}},
  \bibinfo {author} {\bibfnamefont {M.}~\bibnamefont {Amado}}, \bibinfo
  {author} {\bibfnamefont {M.}~\bibnamefont {Orlita}}, \bibinfo {author}
  {\bibfnamefont {M.}~\bibnamefont {Sprinkle}}, \bibinfo {author}
  {\bibfnamefont {C.}~\bibnamefont {Berger}}, \bibinfo {author} {\bibfnamefont
  {W.~A.}\ \bibnamefont {de~Heer}}, \ and\ \bibinfo {author} {\bibfnamefont
  {M.}~\bibnamefont {Potemski}},\ }\href@noop {} {\bibfield  {journal}
  {\bibinfo  {journal} {Phys. Rev. Lett.}\ }\textbf {\bibinfo {volume} {103}},\
  \bibinfo {pages} {186803} (\bibinfo {year} {2009})}\BibitemShut {NoStop}%
\bibitem [{\citenamefont {Yan}\ \emph {et~al.}(2010)\citenamefont {Yan},
  \citenamefont {Goler}, \citenamefont {Rhone}, \citenamefont {Han},
  \citenamefont {He}, \citenamefont {Kim}, \citenamefont {Pellegrini},\ and\
  \citenamefont {Pinczuk}}]{Yan2010}%
  \BibitemOpen
  \bibfield  {author} {\bibinfo {author} {\bibfnamefont {J.}~\bibnamefont
  {Yan}}, \bibinfo {author} {\bibfnamefont {S.}~\bibnamefont {Goler}}, \bibinfo
  {author} {\bibfnamefont {T.~D.}\ \bibnamefont {Rhone}}, \bibinfo {author}
  {\bibfnamefont {M.}~\bibnamefont {Han}}, \bibinfo {author} {\bibfnamefont
  {R.}~\bibnamefont {He}}, \bibinfo {author} {\bibfnamefont {P.}~\bibnamefont
  {Kim}}, \bibinfo {author} {\bibfnamefont {V.}~\bibnamefont {Pellegrini}}, \
  and\ \bibinfo {author} {\bibfnamefont {A.}~\bibnamefont {Pinczuk}},\
  }\href@noop {} {\bibfield  {journal} {\bibinfo  {journal} {Phys. Rev. Lett.}\
  }\textbf {\bibinfo {volume} {105}},\ \bibinfo {pages} {227401} (\bibinfo
  {year} {2010})}\BibitemShut {NoStop}%
\bibitem [{\citenamefont {Garcia-Flores}\ \emph {et~al.}(2009)\citenamefont
  {Garcia-Flores}, \citenamefont {Terashita}, \citenamefont {Granado},\ and\
  \citenamefont {Kopelevich}}]{Garcia09}%
  \BibitemOpen
  \bibfield  {author} {\bibinfo {author} {\bibfnamefont {A.~F.}\ \bibnamefont
  {Garcia-Flores}}, \bibinfo {author} {\bibfnamefont {H.}~\bibnamefont
  {Terashita}}, \bibinfo {author} {\bibfnamefont {E.}~\bibnamefont {Granado}},
  \ and\ \bibinfo {author} {\bibfnamefont {Y.}~\bibnamefont {Kopelevich}},\
  }\href@noop {} {\bibfield  {journal} {\bibinfo  {journal} {Phys. Rev. B}\
  }\textbf {\bibinfo {volume} {79}},\ \bibinfo {pages} {113105} (\bibinfo
  {year} {2009})}\BibitemShut {NoStop}%
\bibitem [{\citenamefont {Slonczewski}\ and\ \citenamefont
  {Weiss}(1958)}]{Slonczewski58}%
  \BibitemOpen
  \bibfield  {author} {\bibinfo {author} {\bibfnamefont {J.~C.}\ \bibnamefont
  {Slonczewski}}\ and\ \bibinfo {author} {\bibfnamefont {P.~R.}\ \bibnamefont
  {Weiss}},\ }\href@noop {} {\bibfield  {journal} {\bibinfo  {journal} {Phys.
  Rev.}\ }\textbf {\bibinfo {volume} {109}},\ \bibinfo {pages} {272} (\bibinfo
  {year} {1958})}\BibitemShut {NoStop}%
\bibitem [{\citenamefont {McClure}(1956)}]{McClure56}%
  \BibitemOpen
  \bibfield  {author} {\bibinfo {author} {\bibfnamefont {J.~W.}\ \bibnamefont
  {McClure}},\ }\href@noop {} {\bibfield  {journal} {\bibinfo  {journal} {Phys.
  Rev.}\ }\textbf {\bibinfo {volume} {104}},\ \bibinfo {pages} {666} (\bibinfo
  {year} {1956})}\BibitemShut {NoStop}%
\bibitem [{\citenamefont {Li}\ \emph {et~al.}(2009{\natexlab{a}})\citenamefont
  {Li}, \citenamefont {Luican},\ and\ \citenamefont {Andrei}}]{Li2009}%
  \BibitemOpen
  \bibfield  {author} {\bibinfo {author} {\bibfnamefont {G.}~\bibnamefont
  {Li}}, \bibinfo {author} {\bibfnamefont {A.}~\bibnamefont {Luican}}, \ and\
  \bibinfo {author} {\bibfnamefont {E.~Y.}\ \bibnamefont {Andrei}},\
  }\href@noop {} {\bibfield  {journal} {\bibinfo  {journal} {Phys. Rev. Lett.}\
  }\textbf {\bibinfo {volume} {102}},\ \bibinfo {pages} {176804} (\bibinfo
  {year} {2009}{\natexlab{a}})}\BibitemShut {NoStop}%
\bibitem [{\citenamefont {Neugebauer}\ \emph {et~al.}(2009)\citenamefont
  {Neugebauer}, \citenamefont {Orlita}, \citenamefont {Faugeras}, \citenamefont
  {Barra},\ and\ \citenamefont {Potemski}}]{Neugebauer2009}%
  \BibitemOpen
  \bibfield  {author} {\bibinfo {author} {\bibfnamefont {P.}~\bibnamefont
  {Neugebauer}}, \bibinfo {author} {\bibfnamefont {M.}~\bibnamefont {Orlita}},
  \bibinfo {author} {\bibfnamefont {C.}~\bibnamefont {Faugeras}}, \bibinfo
  {author} {\bibfnamefont {A.-L.}\ \bibnamefont {Barra}}, \ and\ \bibinfo
  {author} {\bibfnamefont {M.}~\bibnamefont {Potemski}},\ }\href@noop {}
  {\bibfield  {journal} {\bibinfo  {journal} {Phys. Rev. Lett.}\ }\textbf
  {\bibinfo {volume} {103}},\ \bibinfo {pages} {136403} (\bibinfo {year}
  {2009})}\BibitemShut {NoStop}%
\bibitem [{\citenamefont {Nakao}(1976)}]{Nakao1976}%
  \BibitemOpen
  \bibfield  {author} {\bibinfo {author} {\bibfnamefont {K.}~\bibnamefont
  {Nakao}},\ }\href@noop {} {\bibfield  {journal} {\bibinfo  {journal} {J.
  Phys. Soc.Jpn.}\ }\textbf {\bibinfo {volume} {40}},\ \bibinfo {pages} {761}
  (\bibinfo {year} {1976})}\BibitemShut {NoStop}%
\bibitem [{\citenamefont {Soule}(1958)}]{Soule1958}%
  \BibitemOpen
  \bibfield  {author} {\bibinfo {author} {\bibfnamefont {D.~E.}\ \bibnamefont
  {Soule}},\ }\href {\doibase 10.1103/PhysRev.112.698} {\bibfield  {journal}
  {\bibinfo  {journal} {Phys. Rev.}\ }\textbf {\bibinfo {volume} {112}},\
  \bibinfo {pages} {698} (\bibinfo {year} {1958})}\BibitemShut {NoStop}%
\bibitem [{\citenamefont {Soule}\ \emph {et~al.}(1964)\citenamefont {Soule},
  \citenamefont {McClure},\ and\ \citenamefont {Smith}}]{Soule1964}%
  \BibitemOpen
  \bibfield  {author} {\bibinfo {author} {\bibfnamefont {D.~E.}\ \bibnamefont
  {Soule}}, \bibinfo {author} {\bibfnamefont {J.~W.}\ \bibnamefont {McClure}},
  \ and\ \bibinfo {author} {\bibfnamefont {L.~B.}\ \bibnamefont {Smith}},\
  }\href {\doibase 10.1103/PhysRev.134.A453} {\bibfield  {journal} {\bibinfo
  {journal} {Phys. Rev.}\ }\textbf {\bibinfo {volume} {134}},\ \bibinfo {pages}
  {A453} (\bibinfo {year} {1964})}\BibitemShut {NoStop}%
\bibitem [{\citenamefont {Woollam}(1970)}]{Woollam1970}%
  \BibitemOpen
  \bibfield  {author} {\bibinfo {author} {\bibfnamefont {J.~A.}\ \bibnamefont
  {Woollam}},\ }\href {\doibase 10.1103/PhysRevLett.25.810} {\bibfield
  {journal} {\bibinfo  {journal} {Phys. Rev. Lett.}\ }\textbf {\bibinfo
  {volume} {25}},\ \bibinfo {pages} {810} (\bibinfo {year} {1970})}\BibitemShut
  {NoStop}%
\bibitem [{\citenamefont {Schneider}\ \emph {et~al.}(2009)\citenamefont
  {Schneider}, \citenamefont {Orlita}, \citenamefont {Potemski},\ and\
  \citenamefont {Maude}}]{Schneider2009}%
  \BibitemOpen
  \bibfield  {author} {\bibinfo {author} {\bibfnamefont {J.~M.}\ \bibnamefont
  {Schneider}}, \bibinfo {author} {\bibfnamefont {M.}~\bibnamefont {Orlita}},
  \bibinfo {author} {\bibfnamefont {M.}~\bibnamefont {Potemski}}, \ and\
  \bibinfo {author} {\bibfnamefont {D.~K.}\ \bibnamefont {Maude}},\ }\href@noop
  {} {\bibfield  {journal} {\bibinfo  {journal} {Phys. Rev. Lett.}\ }\textbf
  {\bibinfo {volume} {102}},\ \bibinfo {pages} {166403} (\bibinfo {year}
  {2009})}\BibitemShut {NoStop}%
\bibitem [{\citenamefont {Abergel}\ and\ \citenamefont
  {Fal'ko}(2007)}]{Abergel2007}%
  \BibitemOpen
  \bibfield  {author} {\bibinfo {author} {\bibfnamefont {D.}~\bibnamefont
  {Abergel}}\ and\ \bibinfo {author} {\bibfnamefont {V.}~\bibnamefont
  {Fal'ko}},\ }\href@noop {} {\bibfield  {journal} {\bibinfo  {journal} {Phys.
  Rev. B}\ }\textbf {\bibinfo {volume} {75}},\ \bibinfo {pages} {155430}
  (\bibinfo {year} {2007})}\BibitemShut {NoStop}%
\bibitem [{\citenamefont {Partoens}\ and\ \citenamefont
  {Peeters}(2007)}]{Partoens2007}%
  \BibitemOpen
  \bibfield  {author} {\bibinfo {author} {\bibfnamefont {B.}~\bibnamefont
  {Partoens}}\ and\ \bibinfo {author} {\bibfnamefont {F.~M.}\ \bibnamefont
  {Peeters}},\ }\href@noop {} {\bibfield  {journal} {\bibinfo  {journal} {Phys.
  Rev. B}\ }\textbf {\bibinfo {volume} {75}},\ \bibinfo {pages} {193402}
  (\bibinfo {year} {2007})}\BibitemShut {NoStop}%
\bibitem [{\citenamefont {Koshino}\ and\ \citenamefont
  {Ando}(2008)}]{Koshino2008}%
  \BibitemOpen
  \bibfield  {author} {\bibinfo {author} {\bibfnamefont {M.}~\bibnamefont
  {Koshino}}\ and\ \bibinfo {author} {\bibfnamefont {T.}~\bibnamefont {Ando}},\
  }\href@noop {} {\bibfield  {journal} {\bibinfo  {journal} {Phys. Rev. B}\
  }\textbf {\bibinfo {volume} {77}},\ \bibinfo {pages} {115313} (\bibinfo
  {year} {2008})}\BibitemShut {NoStop}%
\bibitem [{\citenamefont {Basko}(2008)}]{Basko2008}%
  \BibitemOpen
  \bibfield  {author} {\bibinfo {author} {\bibfnamefont {D.~M.}\ \bibnamefont
  {Basko}},\ }\href@noop {} {\bibfield  {journal} {\bibinfo  {journal} {Phys.
  Rev. B}\ }\textbf {\bibinfo {volume} {78}},\ \bibinfo {pages} {125418}
  (\bibinfo {year} {2008})}\BibitemShut {NoStop}%
\bibitem [{\citenamefont {Basko}(2009)}]{Basko2009}%
  \BibitemOpen
  \bibfield  {author} {\bibinfo {author} {\bibfnamefont {D.~M.}\ \bibnamefont
  {Basko}},\ }\href@noop {} {\bibfield  {journal} {\bibinfo  {journal} {New. J.
  Phys.}\ }\textbf {\bibinfo {volume} {11}},\ \bibinfo {pages} {095011}
  (\bibinfo {year} {2009})}\BibitemShut {NoStop}%
\bibitem [{\citenamefont {Ando}(2007{\natexlab{b}})}]{Ando2007b}%
  \BibitemOpen
  \bibfield  {author} {\bibinfo {author} {\bibfnamefont {T.}~\bibnamefont
  {Ando}},\ }\href@noop {} {\bibfield  {journal} {\bibinfo  {journal} {J. Phys.
  Soc. Jpn.}\ }\textbf {\bibinfo {volume} {76}},\ \bibinfo {pages} {104711}
  (\bibinfo {year} {2007}{\natexlab{b}})}\BibitemShut {NoStop}%
\bibitem [{\citenamefont {Li}\ \emph {et~al.}(2009{\natexlab{b}})\citenamefont
  {Li}, \citenamefont {Henriksen}, \citenamefont {Jiang}, \citenamefont {Hao},
  \citenamefont {M.C. Martin~a}, \citenamefont {Stormer},\ and\ \citenamefont
  {Basov}}]{Li2009n}%
  \BibitemOpen
  \bibfield  {author} {\bibinfo {author} {\bibfnamefont {Z.}~\bibnamefont
  {Li}}, \bibinfo {author} {\bibfnamefont {E.}~\bibnamefont {Henriksen}},
  \bibinfo {author} {\bibfnamefont {Z.}~\bibnamefont {Jiang}}, \bibinfo
  {author} {\bibfnamefont {Z.}~\bibnamefont {Hao}}, \bibinfo {author}
  {\bibfnamefont {d.~P.~K.}\ \bibnamefont {M.C. Martin~a}}, \bibinfo {author}
  {\bibfnamefont {H.}~\bibnamefont {Stormer}}, \ and\ \bibinfo {author}
  {\bibfnamefont {D.}~\bibnamefont {Basov}},\ }\href@noop {} {\bibfield
  {journal} {\bibinfo  {journal} {Nature Phys.}\ }\textbf {\bibinfo {volume}
  {4}},\ \bibinfo {pages} {532} (\bibinfo {year}
  {2009}{\natexlab{b}})}\BibitemShut {NoStop}%
\bibitem [{\citenamefont {Brandt}\ \emph {et~al.}(1988)\citenamefont {Brandt},
  \citenamefont {Chudinov},\ and\ \citenamefont {Ponomarev}}]{Brandt1988}%
  \BibitemOpen
  \bibfield  {author} {\bibinfo {author} {\bibfnamefont {N.}~\bibnamefont
  {Brandt}}, \bibinfo {author} {\bibfnamefont {S.}~\bibnamefont {Chudinov}}, \
  and\ \bibinfo {author} {\bibfnamefont {Y.}~\bibnamefont {Ponomarev}},\
  }\href@noop {} {\emph {\bibinfo {title} {Semimetals 1: Graphite and its
  Compounds}}},\ edited by\ \bibinfo {editor} {\bibnamefont {North-Holland}}\
  (\bibinfo {year} {1988})\BibitemShut {NoStop}%
\bibitem [{\citenamefont {McCann}\ and\ \citenamefont
  {Fal'ko}(2006)}]{McCann2006}%
  \BibitemOpen
  \bibfield  {author} {\bibinfo {author} {\bibfnamefont {E.}~\bibnamefont
  {McCann}}\ and\ \bibinfo {author} {\bibfnamefont {V.}~\bibnamefont
  {Fal'ko}},\ }\href@noop {} {\bibfield  {journal} {\bibinfo  {journal} {Phys.
  Rev. Lett.}\ }\textbf {\bibinfo {volume} {96}},\ \bibinfo {pages} {086805}
  (\bibinfo {year} {2006})}\BibitemShut {NoStop}%
\end{thebibliography}
\end{document}